\documentclass[12pt]{article}
\usepackage{amscd}
\usepackage{bbm}
\usepackage{mathrsfs}
\usepackage{amssymb}
\usepackage{graphics}
\usepackage{graphicx}
\usepackage{amsfonts}
\usepackage{amsmath}
\usepackage{caption}
\usepackage{titlesec}
\usepackage{subcaption}
\usepackage{float}
\usepackage[numbers,sort&compress]{natbib}
\titleformat{\section}{\large\bfseries}{\thesection. }{0em}{}
\titleformat{\subsection}{\normalsize\bfseries}{\thesubsection. }{0em}{}
\numberwithin{equation}{section}
\begin{document}
	\date{}
	\title{Mixed single, double, and triple poles solutions for the space-time shifted nonlocal DNLS equation with nonzero boundary conditions via Riemann--Hilbert approach}
	
	\author{Xin-Yu Liu, Rui Guo$ \thanks{Corresponding author,
			guorui@tyut.edu.cn}$ \
		\\
		\\{\em
			School of Mathematics, Taiyuan University of} \\
		{\em Technology, Taiyuan 030024, China} } \maketitle
	
	\begin{abstract}		
		In this paper, we investigate the space-time shifted nonlocal derivative nonlinear Schr\"{o}dinger (DNLS) equation under nonzero boundary conditions using the Riemann--Hilbert (RH) approach for the first time. To begin with, in the direct scattering problem, we analyze the analyticity, symmetries, and asymptotic behaviors of the Jost eigenfunctions and scattering matrix functions. Subsequently, we examine the coexistence of $N$-single, $N$-double, and $N$-triple poles in the inverse scattering problem. The corresponding residue conditions, trace formulae, $\theta$ condition, and symmetry relations of the norming constants are obtained. Moreover, we derive the exact expression for the mixed single, double, and triple poles solutions with the reflectionless potentials by solving the relevant RH problem associated with the space-time shifted nonlocal DNLS equation. Furthermore, to further explore the remarkable characteristics of soliton solutions, we graphically illustrate the dynamic behaviors of several representative solutions, such as three-soliton, two-breather, and soliton-breather solutions. Finally, we analyze the effects of shift parameters through graphical simulations.
		
		\vspace{7mm}\noindent\emph{Keywords}: 
		Space-time shifted nonlocal DNLS equation; Riemann--Hilbert approach; Nonzero boundary conditions
	\end{abstract}
	\newpage
\section{Introduction}
\hspace{0.7cm}Since Ablowitz introduced an integrable nonlocal nonlinear Schr\"{o}dinger (NLS) equation in 2013 \cite{g1}, research on nonlocal equations has gained significant attention, particularly in integrable theory \cite{g2,g3,g4,g5,g6,g7,g8,g9}. 
Following this, in 2021, Ablowitz further proposed the shifted PT symmetric nonlocal NLS equation
\begin{equation}\label{e4}
	iq_t\left( x,t \right) =q_{xx}\left( x,t \right) +2q^2\left( x,t \right) q^*\left( x_0-x,t \right),
\end{equation}
and derived the soliton solutions by employing the inverse scattering transform \cite{g10}. The three types of Darboux transformations for Eq.~(\ref{e4}) were explored in \cite{g11}. Furthermore, the soliton dynamical behaviors of the shifted nonlocal reverse time NLS equation and the shifted nonlocal reverse space-time NLS equation were investigated in \cite{g12}. Additionally, inverse scattering transform and soliton solutions for both continuous and discrete shifted nonlocal NLS equations and mKdV equations were presented in \cite{g13}. 

The derivative nonlinear Schr\"{o}dinger (DNLS) equation, which is a fundamental nonlinear physical model, has extensive physical applications, such as the propagation of Alfvén waves in plasma physics \cite{g14,g15} and weak nonlinear electromagnetic waves in ferromagnetic \cite{g16,g17}. By applying different nonlocal symmetry reduction, different nonlocal DNLS equations were proposed and studied by Darboux transformation \cite{g18,g19,g20}, inverse scatting transform \cite{g21,g22}, and Riemann--Hilbert (RH) approach \cite{g23}. Subsequently, by imposing the space-time shifted nonlocal integrable symmetry reduction $r(x,t)=\sigma q(x_{0}-x,t_{0}-t)$ in standard DNLS equation, it is natural to obtain the space-time shifted nonlocal DNLS equation
	\begin{equation}\label{e1}
	q_t\left( x,t \right) =iq_{xx}\left( x,t \right) +\sigma \left( q^2\left( x,t \right) q\left( x_0-x,t_0-t \right) \right) _x,
\end{equation}
where $\sigma =\pm 1$, $x_{0}$, $t_{0}$ are arbitrary real constants representing the space and time shifts, respectively. When $x_{0}=0,\ t_{0}=0$, Eq.~(\ref{e1}) simplifies to the nonlocal reverse space-time DNLS equation, which we investigated soliton solutions for the single-pole and double-pole cases in \cite{g24}. To our knowledge, extensive research has been conducted on higher-order soliton solutions in both local \cite{g25,g26,g27,g28,g29,g30} and nonlocal \cite{g31,g32,g33} equations, but the higher-order soliton solutions of Eq.~(\ref{e1}) are still underexplored, mainly due to the increased complexity of deriving the symmetry relations of scattering data in the shifted nonlocal case. 

In this paper, we utilize the RH approach to investigate the Eq.~(\ref{e1}) under nonzero boundary conditions (NZBCs)
\begin{equation}\label{e6}
	q\left( x,t \right) \rightarrow q_{\pm},\ x\rightarrow \pm \infty ,\\
\end{equation}
where\ $ |q_{\pm}|=q_{0}>0 $ and $q_{\pm}=\eta q_{\mp}^{\ast},\ \eta =\pm 1$. This is the first time the RH approach has been extended to the shifted nonlocal DNLS equations. It is worth noting that our study does not focus on a simple scenario involving only one type of pole, such as $N$-single poles or $N$-double poles; rather, we explore a mixed situation where single, double, and triple poles coexist, which is significantly more complex. Moreover, under certain parameter conditions, we apply the derived exact expression for the mixed single, double, and triple poles solutions to simulate the dynamics of three-soliton, two-breather, soliton-breather solutions, and more. The roles of shift parameters can be seen from the dynamic analysis of the graphs. To our knowledge, these results have not been reported.

The structure of this paper is as follows: In Section 2, the properties of Jost eigenfunctions and scattering matrix functions under NZBCs (\ref{e6}) are presented. In Section 3, we construct the RH problem, analyze the distribution of discrete eigenvalues, and give the trace formulae and $\theta $ condition. In Section 4, the exact expression for the mixed single, double, and triple poles solutions with the reflectionless potentials is given. Additionally, the dynamic behaviors of soliton solutions and the effects of shift parameters are illustrated graphically. Section 5 contains the conclusions.
\section{Direct Scattering problem}
\hspace{0.7cm}The direct scattering problem with NZBCs for both the shifted nonlocal reverse space-time DNLS equation and the unshifted nonlocal cases are similar. For brevity, we directly present some results for the direct scattering problem. The Eq.~(\ref{e1}) admits the Lax pair \cite{g8}
\begin{align}\label{e2}
	\varPsi _x&=X\varPsi ,\ X\left( x,t,\lambda \right) =-i\lambda ^2\sigma _3+\lambda Q ,\\\label{e3}
	\varPsi _t&=T\varPsi ,\ T\left( x,t,\lambda \right) =-2i\lambda ^4\sigma _3+2\lambda ^3Q-i\lambda ^2Q^2-i\lambda Q_x+\lambda Q^3 , 
\end{align}
where 
	\begin{align}
		\sigma _3=\left( \begin{matrix}
			1&		0\\
			0&		-1\\
		\end{matrix} \right),\ 
		Q=\left( \begin{matrix}
			0&		q(x,t)\\
			\sigma q(x_0-x,t_0-t)&		0\\
		\end{matrix} \right) . \nonumber
	\end{align}
\hspace{0.7cm}When $x\rightarrow \pm \infty$, the asymptotic scattering problem can be given by
 \begin{align}\label{e21}
 	\varPsi _x&=X_{\pm}\varPsi ,\   X_{\pm}\left( x,t,\lambda \right) =-i\lambda ^2\sigma _3+\lambda Q_{\pm},\\\label{e22}
 	\varPsi _t&=T_{\pm}\varPsi ,\ \  T_{\pm}\left( x,t,\lambda \right) =\left( 2\lambda ^2+\sigma \eta q_0^2 \right) X_{\pm},
 \end{align}
 where
 $$ 
 Q_{\pm}=\left( \begin{matrix}
 	0&		q_{\pm}\\
 	\sigma q_{\mp}&		0\\
 \end{matrix} \right) =\left( \begin{matrix}
 	0&		q_{\pm}\\
 	\sigma \eta q_{\pm}^{\ast}&		0\\
 \end{matrix} \right).
 $$
 
 By introducing a two-sheeted Riemann surface $k^2=\lambda^2-\sigma\eta q_{0}^{2}$ and the uniformization variable $z=k+\lambda $, we can analyze the scattering problem in the complex $z$ plane. For convenience, let $D_{\pm}=\left\{ z\in \mathbb{C}|\pm \text{Re}z\text{Im}z>0 \right\},\ \Sigma =\mathbb{R}\cup i\mathbb{R}\setminus \left\{ 0 \right\}$. Then, the solutions of the Lax pair (\ref{e21})-(\ref{e22}) under NZBCs (\ref{e6}) can be derived as follows:
   	\begin{equation}
		\varPsi _{\pm}\sim  M_{\pm}e^{i\theta \left( z \right) \sigma _3}, \   x\rightarrow \pm \infty , 
	\end{equation}
	where $\theta \left( z \right) =k\left( z \right) \lambda \left( z \right) \left[ x+\left( 2 \lambda^2\left( z \right) +\sigma \eta q_{0}^{2} \right) t \right]$, $k\left( z \right)=\frac{1}{2}\left( z-\sigma \eta \frac{q_0^2}{z} \right),\ \lambda\left( z \right) =\frac{1}{2}\left( z+\sigma \eta \frac{q_0^2}{z} \right)$, $M_{\pm}=\left( \begin{matrix}
		\frac{-iq_{\pm}}{z}&		1\\
		1&		\frac{i\sigma \eta q_{\pm}^{\ast}}{z}\\
	\end{matrix} \right). $
	
Then, the modified Jost eigenfunctions $\nu _{\pm}\left( x,t,z \right) $ are obtained by the transform $\nu _{\pm}=\varPsi _{\pm}e^{-i\theta\left( z \right)  \sigma _3}$ such that $\nu _{\pm}\rightarrow M_{\pm}$ as $ x\rightarrow \pm \infty $. It is evident from Eq.~(\ref{e2}) that the modified Jost eigenfunctions $\nu _{\pm}\left( x,t,z \right) $ satisfy the following Volterra integral equations:
	\begin{equation}\label{e16}
		  \nu _{\pm}\left( x,t,z \right) =M _{\pm}+\lambda \int_{\pm\infty}^x{M _{\pm}e}^{ik\lambda \left( x-y \right) \hat{\sigma}_3}\left[ M _{\pm}^{-1}\varDelta Q _{\pm}\left( y,t,z \right) \nu _{\pm}\left( y,t,z \right) \right] dy,
	\end{equation}
where $e^{i\alpha \widehat{\sigma _3}}X=e^{i\alpha \sigma _3}Xe^{-i\alpha \sigma _3}$.

	Because the Jost eigenfunctions $\varPsi _{\pm}\left( x,t,z \right) $ are the fundamental solutions of Eqs.~(\ref{e2})-(\ref{e3}), there exists a constant scattering matrix $ U(z) $ such that
	\begin{equation}\label{e17}
		\nu _+\left( x,t,z \right) =\nu _-\left( x,t,z \right) e^{i\theta\left( z \right)  \hat{\sigma}_3}U\left( z \right) ,
	\end{equation}
	where
	$$
	U\left( z \right) =\left( \begin{matrix}
		u_{11}\left( z \right)&		u_{12}\left( z \right)\\
		u_{21}\left( z \right)&		u_{22}\left( z \right)\\
	\end{matrix} \right) .
	$$
	
	Furthermore, one has the following propositions.
	
	\textbf{Proposition 1.} Assume $q-q_{\pm}\in L^1\left( \mathbb{R} \right) $, $\nu _{+,1}$, $\nu _{-,2}$, and $u_{11}(z)$ can be analytically extended to $D_{+}$, while $\nu _{+,2}$, $\nu _{-,1}$, and $u_{22}(z)$ can be analytically extended to $D_{-}$, $u_{12}(z)$ and $u_{21}(z)$ are continuous in $\Sigma $. 
	
	\textit{\textbf{Proof}} In terms of Volterra integral equations (\ref{e16}), it is easy to prove the existence, uniqueness, and analyticity of $\nu _{\pm}$ by applying mathematical induction. Then, based on the analyticity of $\nu _{\pm}$ and the relation (\ref{e17}), the analyticity of the scattering coefficient $u_{ij},\ i,j=1,2$ can obviously be deduced.$\hfill\square$
	
	\textbf{Proposition 2.} The symmetries of the Jost eigenfunctions and the scattering matrix functions are as follows:
	\begin{align}
		&\varPsi _{\pm}\left( x,t,z \right)=\left\{ \begin{array}{l}
			\sigma _1\varPsi _{\mp}\left( x_0-x,t_0-t,-z \right) \sigma _1e^{i\theta _0\left( -z \right) \sigma _3},\,\sigma =1,\\
			-\sigma _2\varPsi _{\mp}\left( x_0-x,t_0-t,-z \right) \sigma _2e^{i\theta _0\left( -z \right) \sigma _3},\,\sigma =-1,\\
		\end{array} \right.\\
		&\varPsi _{\pm}\left( x,t,z \right) =\frac{i}{z}\varPsi _{\pm}\left( x,t,\sigma \eta \frac{q_{0}^{2}}{z} \right) \sigma _1Q_{\pm}\sigma _{\ast} ,\ \varPsi _{\pm}\left( x,t,z \right) =-\sigma _3\varPsi _{\pm}\left( x,t,-z \right) \sigma _3 , \\ \label{e7}
		&U\left( z \right) =\left\{ \begin{array}{l}
			\sigma _1\left( U\left( -z \right) \right) ^{-1}\sigma _1e^{i\theta _0\left( -z \right)\sigma _3},\ \sigma =1 ,\\
			\sigma _2\left( U\left( -z \right) \right) ^{-1}\sigma _2e^{i\theta _0\left( -z \right)\sigma _3},\ \sigma =-1,\\
		\end{array} \right.\\ \label{e9}
		&U\left( z \right) =\left( \sigma _1Q_-\sigma _{\ast} \right) ^{-1}U\left( \sigma \eta \frac{q_{0}^{2}}{z} \right) \sigma _1Q_+\sigma _{\ast},\ U\left( z \right) =\sigma _3U\left( -z \right) \sigma _3,
	\end{align}
	where $\theta _0\left( z \right) =k\left( z \right) \lambda \left( z \right) \left( x_0+\left( 2\lambda ^2\left( z \right) +\sigma \eta q_{0}^{2} \right) t_0 \right)$, and $$\sigma _1=\left( \begin{matrix}
			0&		1\\
			1&		0\\
		\end{matrix} \right) ,\ \sigma _2=\left( \begin{matrix}
			0&		-i\\
			i&		0\\
		\end{matrix} \right) ,\ \sigma _{\ast}=\left( \begin{matrix}
			0&		1\\
			-1&		0\\
		\end{matrix} \right).$$		
		
\textbf{Proposition 3.} The asymptotic properties of the Jost eigenfunctions and scattering matrix functions are as follows:
	\begin{align}
		&z\rightarrow \infty:\ \nu _{\pm}\left( x,t,z \right) =e^{im_{\pm}\sigma _{\ast}}+O\left( \frac{1}{z} \right) ,\ U\left( z \right) =e^{i\bar{m}\sigma _3}+O\left( \frac{1}{z} \right) ,\\
        &z\rightarrow 0:\ \nu _{\pm}\left( x,t,z \right) =\frac{i\sigma \eta \sigma _3e^{im_{\pm}\sigma _{\ast}}Q_{\pm}^{\ast}}{z}+O\left( 1 \right) ,\ U\left( z \right) =\text{diag}\left( \frac{q_+}{q_-},\frac{q_-}{q_+} \right) e^{-i\bar{m}\sigma _3}+O\left( z \right) ,
	\end{align}
	where
	$$
	m_{\pm}=-\frac{1}{2}\int_{\pm \infty}^x{\left( \sigma \eta q_{0}^{2}-\sigma q\left( y,t \right) q\left( x_0-y,t_0-t \right) \right) dy},
	$$
	$$
	 \bar{m}=\frac{1}{2}\int_{-\infty}^{\infty}{\left( \sigma q\left( y,t \right) q\left( x_0-y,t_0-t \right) -\sigma \eta q_0^2 \right)}dy . 
    $$
	\section{Inverse scattering problem}
	\hspace{0.7cm}By introducing the sectionally meromorphic matrices
	\begin{equation}\label{e18}
		W\left( x,t,z \right) =\left\{ \begin{array}{l}
			W^+\left( x,t,z \right) =\left( \frac{\nu _{+,1}\left( x,t,z \right)}{u_{11}\left( z \right)},\nu _{-,2}\left( x,t,z \right) \right) ,\ z\in D_+,\\
			W^-\left( x,t,z \right) =\left( \nu _{-,1}\left( x,t,z \right) ,\frac{\nu _{+,2}\left( x,t,z \right)}{u_{22}\left( z \right)} \right) ,\ z\in D_-,\\
		\end{array} \right. 
	\end{equation}
    the following RH problem is derived:\\
	(1) $W^\pm$ are meromorphic in $D_{\pm}\setminus \Sigma$ respectively, \\
	(2) $W^-\left( x,t,z \right) =W^+\left( x,t,z \right) \left( I-J\left( x,t,z \right) \right)$, where $$
	J=e^{i\theta \left( z \right) \hat{\sigma}_3}\left( \begin{matrix}
		0&		-\tilde{\rho}\left( z \right)\\
		\rho \left( z \right)&		\rho \left( z \right) \tilde{\rho}\left( z \right)\\
	\end{matrix} \right),\ \rho \left( z \right)=\frac{u_{21}(z)}{u_{11}(z)},\ \tilde{\rho}\left( z \right) =\frac{u_{12}(z)}{u_{22}(z)},$$
	(3) $W^{\pm}\left( x,t,z \right) =e^{im_-\sigma _{\ast}}+O\left( \frac{1}{z} \right)\ $as $z\rightarrow \infty$, $W^{\pm}\left( x,t,z \right) =\frac{i}{z}\sigma \eta e^{im_-\sigma _{\ast}}\sigma _3Q_-^{\ast}+O\left( 1 \right)$ as $z\rightarrow 0.$
	
Substituting $\nu _{\pm}=\varPsi _{\pm}e^{-i\theta\left( z \right)  \sigma _3}$ into Eq.~(\ref{e2}), one can compare the power coefficient of $z$ and combine the definition (\ref{e18}), then the solution of Eqs.~(\ref{e1})-(\ref{e6}) can be expressed by the solution of RH problem:
	\begin{equation}\label{e15}
		q\left( x,t \right) =ie^{im_{\pm}}\underset{z\rightarrow \infty}{\lim}\left( z\nu _{\pm} \right) _{11}=ie^{im_-}\underset{z\rightarrow \infty}{\lim}\left( zW^- \right) _{11} .
	\end{equation}

	Then, we suppose that $u_{11} (z) $ has $N_1$ single zeros $\xi_{1,n}$, $N_2$ double zeros $\xi_{2,n}$, and $N_3$ triple zeros $\xi_{3,n}$ in $D _{+}\bigcap\left\{ z\in \mathbb{C}\mid \text{Re}z>0,\text{Im}z>0 \right\}$, then $u_{11}\left( \xi _{1,n} \right) =0,\ u_{11}^{'}\left( \xi _{1,n} \right) \ne 0,\ u_{11}\left( \xi _{2,n} \right) =u_{11}^{'}\left( \xi _{2,n} \right) =0,\ u_{11}^{''}\left( \xi _{2,n} \right) \ne 0,\ u_{11}\left( \xi _{3,n} \right) =u_{11}^{'}\left( \xi _{3,n} \right) =u_{11}^{''}\left( \xi _{3,n} \right) = 0,\ u_{11}^{'''}\left( \xi _{3,n} \right) \ne 0$. Based on the symmetries (\ref{e7})-(\ref{e9}), it can be inferred that $u_{11}\left(z \right) =u_{11}\left( -z \right)=u_{22}\left( \frac{\sigma \eta q_{0}^{2}}{z} \right) =u_{22}\left( -\frac{\sigma \eta q_{0}^{2}}{z} \right)$. For simplicity, we define\\
 \(\xi_n =
\begin{cases}
	\xi_{1,n}, & n = 1, \, 2, \, \cdots, \, N_1, \\
	-\xi_{1,n - N_1}, & n = N_1 + 1, \, N_1 + 2, \, \cdots, \, 2N_1, \\
	\xi_{2,n - 2N_1}, & n = 2N_1 + 1, \, 2N_1 + 2, \, \cdots, \, 2N_1 + N_2, \\
	-\xi_{2,n - 2N_1 - N_2}, & n = 2N_1 + N_2 + 1, \, 2N_1 + N_2 + 2, \, \cdots, \, 2N_1 + 2N_2, \\
	\xi_{3,n - 2N_1 - 2N_2}, & n = 2N_1 + 2N_2 + 1, \, 2N_1 + 2N_2 + 2, \, \cdots, \, 2N_1 + 2N_2 + N_3, \\
	-\xi_{3,n - 2N_1 - 2N_2 - N_3}, & n = 2N_1 + 2N_2 + N_3 + 1, \, 2N_1 + 2N_2 + N_3 + 2, \, \cdots, \, 2N_1 + 2N_2 + 2N_3,
\end{cases}\)
	and $\widehat{\xi }_n=\frac{\sigma \eta q_{0}^{2}}{\xi _n}$, so we have $4N_1+4N_2+4N_3 $ discrete eigenvalues, and the discrete spectrum is given by $Z=\left\{ \xi _n,\ \widehat{\xi }_n,\ n=1,\cdot \cdot \cdot ,\ 2N_1+2N_2+2N_3,\ n\in \mathbb{N}\right\} $. Then for a given $\xi _n\in Z\cap D_+$, we introduce the norming constants $b\left( \xi _n\right)$, $d\left( \xi _n \right)$, and $h\left( \xi _n \right)$ such that
	\begin{equation}
		\begin{aligned}\label{e19}
			&\varPsi _{+,1}\left( \xi _n \right) =b\left( \xi _n \right) \varPsi _{-,2}\left( \xi _n \right) ,\ n\in \left[1,2L\right],\ L=N_1+N_2+N_3,\\
            &\varPsi _{+,1}^{'}\left( \xi _n \right) =b\left( \xi _n \right) \varPsi _{-,2}^{'}\left( \xi _n \right) +d\left( \xi _n \right) \varPsi _{-,2}\left( \xi _n \right) ,\ n\in \left[2N_1+1,2L\right],\\
			&\varPsi _{+,1}^{''}\left( \xi _n \right) =b\left( \xi _n \right) \varPsi _{-,2}^{''}\left( \xi _n \right) +2d\left( \xi _n \right) \varPsi _{-,2}^{'}\left( \xi _n \right) +h\left( \xi _n \right) \varPsi _{-,2}\left( \xi _n \right),\ n\in \left[2N_1+2N_2+1,2L\right].
		\end{aligned}
	\end{equation}
	Similarly, for a given $ \widehat{\xi _n}\in Z\cap D_-$, one can obtain that
	\begin{equation}
		\begin{aligned}\label{e20}
			&\varPsi _{+,2}\left( \widehat{\xi _n} \right) =b\left( \widehat{\xi _n} \right) \varPsi _{-,1}\left( \widehat{\xi _n} \right) ,\ n\in \left[1,2L\right],\\
			&\varPsi _{+,2}^{'}\left( \widehat{\xi _n} \right) =b\left( \widehat{\xi _n} \right) \varPsi _{-,1}^{'}\left( \widehat{\xi _n} \right) +d\left( \widehat{\xi _n} \right) \varPsi _{-,1}\left( \widehat{\xi _n} \right) ,\ n\in \left[2N_1+1,2L\right],\\
			&\varPsi _{+,2}^{''}\left( \widehat{\xi _n} \right) =b\left( \xi _n \right) \varPsi _{-,1}^{''}\left( \widehat{\xi _n} \right) +2d\left( \widehat{\xi _n} \right) \varPsi _{-,1}^{'}\left( \widehat{\xi _n} \right) +h\left( \widehat{\xi _n} \right) \varPsi _{-,1}\left( \widehat{\xi _n} \right) ,\ n\in \left[2N_1+2N_2+1,2L\right],
		\end{aligned}
	\end{equation}
	where $b\left(\widehat{\xi _n}\right)$, $d\left( \widehat{\xi _n} \right)$, and $h\left( \widehat{\xi _n} \right)$ are constants independent of $x$ and $t$.
	
	Furthermore, the pole contributions with single, double, and triple poles of $W^+$ and $W^-$ can be taken as:\\ 
	(1) single poles:
\begin{equation*}	
	\underset{z=\xi _n}{\text{Res}}\,\,W^+\left( z \right) =\left( A\left( \xi _n \right) \nu _{-,2}\left( \xi _n \right) ,0 \right) ,\ \underset{z=\widehat{\xi _n}}{\text{Res}}\,\,W^-\left( z \right) =\left( 0,A\left( \widehat{\xi _n} \right) \nu _{-,1}\left( \widehat{\xi _n} \right) \right) ,\,\,
\end{equation*}
where
$$
A\left( \xi _n \right) =\frac{b\left( \xi _n \right) e^{-2i\theta \left( \xi _n \right)}}{u_{11}^{'}\left( \xi _n \right)},\ A\left( \widehat{\xi _n} \right) =\frac{b\left( \widehat{\xi _n} \right) e^{2i\theta \left( \widehat{\xi _n} \right)}}{u_{22}^{'}\left( \widehat{\xi _n} \right)}.
$$
	(2) double poles:
\begin{equation*}
\begin{aligned}
	&\underset{z=\xi _n}{P_{-2}}\,\,W^+\left( z \right) =\left( B\left( \xi _n \right) \nu _{-,2}\left( \xi _n \right) ,0 \right) ,\ \underset{z=\widehat{\xi _n}}{P_{-2}}\,\,W^-\left( z \right) =\left( 0,B\left( \widehat{\xi _n} \right) \nu _{-,1}\left( \widehat{\xi _n} \right) \right) ,\\
	&	\underset{z=\xi _n}{\text{Res}}\,\,W^+\left( z \right) =\left( B\left( \xi _n \right) C\left( \xi _n \right) \nu _{-,2}\left( \xi _n \right) +B\left( \xi _n \right) \nu _{-,2}^{'}\left( \xi _n \right) ,0 \right) ,\\
	& \underset{z=\widehat{\xi _n}}{\text{Res}}\,\,W^-\left( z \right) =\left( 0,B\left( \widehat{\xi _n} \right) C\left( \widehat{\xi _n} \right) \nu _{-,1}\left( \widehat{\xi _n} \right) +B\left( \widehat{\xi _n} \right) \nu _{-,1}^{'}\left( \widehat{\xi _n} \right) \right) ,
\end{aligned}
\end{equation*}
where
\begin{equation*}
	\begin{aligned}
		&B\left( \xi _n \right) =\frac{2b\left( \xi _n \right) e^{-2i\theta \left( \xi _n \right)}}{u_{11}^{''}\left( \xi _n \right)},\ B\left( \widehat{\xi _n} \right) =\frac{2b\left( \widehat{\xi _n} \right) e^{2i\theta \left( \widehat{\xi _n} \right)}}{u_{22}^{''}\left( \widehat{\xi _n} \right)},\\
		&C\left( \xi _n \right) =\frac{d\left( \xi _n \right)}{b\left( \xi _n \right)}-2i\theta ^{'}\left( \xi _n \right) -\frac{u_{11}^{'''}\left( \xi _n \right)}{3u_{11}^{''}\left( \xi _n \right)},\ C\left( \widehat{\xi _n} \right) =\frac{d\left( \widehat{\xi _n} \right)}{b\left( \widehat{\xi _n} \right)}+2i\theta ^{'}\left( \widehat{\xi _n} \right) -\frac{u_{22}^{'''}\left( \widehat{\xi _n} \right)}{3u_{22}^{''}\left( \widehat{\xi _n} \right)}.
	\end{aligned}	
\end{equation*}
	(3) triple poles:
	\begin{align}
		&	\underset{z=\xi _n}{P_{-3}}\,\,W^+\left( z \right) =\left( D\left( \xi _n \right) \nu _{-,2}\left( \xi _n \right) ,0 \right) ,\ \underset{z=\widehat{\xi _n}}{P_{-3}}\,\,W^-\left( z \right) =\left( 0,D\left( \widehat{\xi _n} \right) \nu _{-,1}\left( \widehat{\xi _n} \right) \right) , \notag\\
		&	\underset{z=\xi _n}{P_{-2}}\,\,W^+\left( z \right) =\left( D\left( \xi _n \right) E\left( \xi _n \right) \nu _{-,2}\left( \xi _n \right) +D\left( \xi _n \right) \nu _{-,2}^{'}\left( \xi _n \right) ,0 \right) , \notag\\
		& \underset{z=\widehat{\xi _n}}{P_{-2}}\,\,W^-\left( z \right) =\left( 0,D\left( \widehat{\xi _n} \right) E\left( \widehat{\xi _n} \right) \nu _{-,1}\left( \widehat{\xi _n} \right) +D\left( \widehat{\xi _n} \right) \nu _{-,1}^{'}\left( \widehat{\xi _n} \right) \right) , \notag\\
		&	\underset{z=\xi _n}{\text{Res}}\,\,W^+\left( z \right) =\left( E\left( \xi _n \right) F\left( \xi _n \right) \nu _{-,2}\left( \xi _n \right) +D\left( \xi _n \right) E\left( \xi _n \right) \nu _{-,2}^{'}\left( \xi _n \right) +\frac{D\left( \xi _n \right)\nu _{-,2}^{''}\left( \xi _n \right)}{2} ,0 \right) , \notag\\
		&\underset{z=\widehat{\xi }_n}{\text{Res}}\,\,W^-\left( z \right) =\left( 0,E\left( \widehat{\xi _n} \right) F\left( \widehat{\xi _n} \right) \nu _{-,1}\left( \widehat{\xi _n} \right) +D\left( \widehat{\xi _n} \right) E\left( \widehat{\xi _n} \right) \nu _{-,1}^{'}\left( \widehat{\xi _n} \right) +\frac{D\left( \widehat{\xi _n} \right)\nu _{-,1}^{''}\left( \widehat{\xi _n} \right) }{2}\right),\notag 
	\end{align}
where
$$
\begin{aligned}
D\left( \xi _n \right) &=\frac{6b\left( \xi _n \right) e^{-2i\theta \left( \xi _n \right)}}{u_{11}^{'''}\left( \xi _n \right)},\
 D\left( \widehat{\xi _n} \right) =\frac{6b\left( \widehat{\xi _n} \right) e^{2i\theta \left( \widehat{\xi _n} \right)}}{u_{22}^{'''}\left( \widehat{\xi _n} \right)},\\
E\left( \xi _n \right) &=\frac{d\left( \xi _n \right)}{b\left( \xi _n \right)}-2i\theta ^{'}\left( \xi _n \right) -\frac{u_{11}^{\left( 4 \right)}\left( \xi _n \right)}{4u_{11}^{'''}\left( \xi _n \right)},\ E\left( \widehat{\xi _n} \right) =\frac{d\left( \widehat{\xi _n} \right)}{b\left( \widehat{\xi _n} \right)}+2i\theta ^{'}\left( \widehat{\xi _n} \right) -\frac{u_{22}^{\left( 4 \right)}\left( \widehat{\xi _n} \right)}{4u_{22}^{'''}\left( \widehat{\xi _n} \right)},\\
F\left( \xi _n \right) &=\frac{h\left( \xi _n \right)}{2b\left( \xi _n \right)}-\frac{d\left( \xi _n \right) u_{11}^{\left( 4 \right)}\left( \xi _n \right)}{4b\left( \xi _n \right) u_{11}^{'''}\left( \xi _n \right)}+\frac{\left( u_{11}^{\left( 4 \right)}\left( \xi _n \right) \right) ^2}{16\left( u_{11}^{'''}\left( \xi _n \right) \right) ^2}-\frac{u_{11}^{\left( 5 \right)}\left( \xi _n \right)}{20u_{11}^{'''}\left( \xi _n \right)}+2\left( \theta ^{'}\left( \xi _n \right) \right) ^2\\
&-i\theta ^{''}\left( \xi _n \right) -2i\theta ^{'}\left( \xi _n \right) E\left( \xi _n \right) ,\\
F\left( \widehat{\xi _n} \right) &=\frac{h\left( \widehat{\xi _n} \right)}{2b\left( \widehat{\xi _n} \right)}-\frac{d\left( \widehat{\xi _n} \right) u_{22}^{\left( 4 \right)}\left( \widehat{\xi _n} \right)}{4b\left( \widehat{\xi _n} \right) u_{22}^{'''}\left( \widehat{\xi _n} \right)}+\frac{\left( u_{22}^{\left( 4 \right)}\left( \widehat{\xi _n} \right) \right) ^2}{16\left( u_{22}^{'''}\left( \widehat{\xi _n} \right) \right) ^2}-\frac{u_{22}^{\left( 5 \right)}\left( \widehat{\xi _n} \right)}{20u_{22}^{'''}\left( \widehat{\xi _n} \right)}+2\left( \theta ^{'}\left( \widehat{\xi _n} \right) \right) ^2 \\
&+i\theta ^{''}\left( \widehat{\xi _n} \right)+2i\theta ^{'}\left( \widehat{\xi _n} \right) E\left( \widehat{\xi _n} \right).
	\end{aligned}
$$

\textbf{Proposition 4.} For $z\in Z$, there are two symmetry relations for $b\left( z \right)$, $d\left( z \right)$, and $h\left( z \right)$: \\
$\bullet$ The first symmetry relation:
$$
\begin{aligned}
	b\left( x,t,z \right) &=\frac{\sigma e^{2i\theta _0\left( -z \right)}}{b\left( x_0-x,t_0-t,-z \right)},\,\,\\
	d\left( x,t,z \right) &=\sigma b^{-2}\left( x_0-x,t_0-t,-z \right) d\left( x_0-x,t_0-t,-z \right) e^{2i\theta _0\left( -z \right)}\\
	&+2i\theta _{0}^{'}\left(-z \right) b^{-1}\left( x_0-x,t_0-t,-z \right) e^{2i\theta _0\left(-z \right)},\\
	h\left( x,t,z \right) &=-\sigma e^{2i\theta _0\left( -z \right)} b^{-2}\left( x_0-x,t_0-t,-z \right) h\left( x_0-x,t_0-t,-z \right)\\
	&+\sigma e^{2i\theta _0\left( -z \right)} b^{-1}\left( x_0-x,t_0-t,-z \right)\left[ 2b^{-2}\left( x_0-x,t_0-t,-z \right) d^2\left( x_0-x,t_0-t,-z \right)\notag\right.\\
	&\left. -4\theta _{0}^{' 2}\left( -z \right)+2i\theta _{0}^{''}\left( -z \right) -4i\theta _{0}^{'}\left( -z \right) b^{-1}\left( x_0-x,t_0-t,-z \right) d\left( x_0-x,t_0-t,-z \right) \right] .\\
\end{aligned}
$$
$\bullet$ The second symmetry relation:
$$
\begin{aligned}
	b\left( z \right) &=-\frac{\sigma}{b\left( \frac{\sigma \eta q_{0}^{2}}{z} \right)},\,\,d\left( z \right) =\frac{\eta q_{0}^{2}}{z^2}d\left( \frac{\sigma \eta q_{0}^{2}}{z} \right) ,\ h\left( z \right) =\frac{-\sigma q_{0}^{4}}{z^4}h\left( \frac{\sigma \eta q_{0}^{2}}{z} \right) -\frac{2\eta q_{0}^{2}}{z^3}d\left( \frac{\sigma \eta q_{0}^{2}}{z} \right) .\\
\end{aligned}
$$

\textit{\textbf{Proof}} Based on Eqs.~(\ref{e19})-(\ref{e20}) and Proposition 2, we can easily derive the above symmetry relations of $b\left( z \right)$, $d\left( z \right)$, and $h\left( z \right)$.$\hfill\square$

According to the definition of discrete spectrum and Plemelj formula, the trace formulae in the case of reflectionless potentials are derived:
	\begin{align}
	&u_{11}\left( z \right) =\prod_{n=1}^{N_1}{\frac{z^2-\xi _{n}^{2}}{z^2-\widehat{\xi _n}^2}}\prod_{n=N_1+1}^{N_1+N_2}{\frac{\left( z^2-\xi _{n}^{2} \right) ^2}{\left( z^2-\widehat{\xi _n}^2 \right) ^2}}\prod_{n=N_1+N_2+1}^{N_1+N_2+N_3}{\frac{\left( z^2-\xi _{n}^{2} \right) ^3}{\left( z^2-\widehat{\xi _n}^2 \right) ^3}}e^{i\bar{m}},\,\,z\in D_+,\\
	&u_{22}\left( z \right) =\prod_{n=1}^{N_1}{\frac{z^2-\widehat{\xi _n}^2}{z^2-\xi _{n}^{2}}}\prod_{n=N_1+1}^{N_1+N_2}{\frac{\left( z^2-\widehat{\xi _n}^2 \right) ^2}{\left( z^2-\xi _{n}^{2} \right) ^2}}\prod_{n=N_1+N_2+1}^{N_1+N_2+N_3}{\frac{\left( z^2-\widehat{\xi _n}^2 \right) ^3}{\left( z^2-\xi _{n}^{2} \right) ^3}}e^{-i\bar{m}},\,\,z\in D_-.
	\end{align}
	When $z\rightarrow 0$, it leads to the $\theta $ condition in the case of reflectionless potentials:	
	\begin{equation}\label{e5}
		\frac{q_+}{q_-}=\prod_{n=1}^{N_1}{\frac{\xi _n^2}{\widehat{\xi _n}^2}}\prod_{n=N_1+1}^{N_1+N_2}{\frac{\xi _n^4}{\widehat{\xi _n}^4}}\prod_{n=N_1+N_2+1}^{N_1+N_2+N_3}{\frac{\xi _n^6}{\widehat{\xi _n}^6}}e^{2i\bar{m}}.
	\end{equation}
	\section{Soliton solutions}
	\hspace{0.7cm}In this section, we will solve the RH problem and derive the exact expression for the solutions of mixed single, double, and triple poles with the reflectionless potentials. Additionally, we will graphically explore the dynamic features of the soliton solutions.
	
	First, to solve the RH problem, we subtract the asymptotic behaviors and the singularity contributions. According to the Plemelj formula, we have
		\begin{equation}\label{e10}
			\begin{aligned}
				W\left( x,t,z \right) &=e^{im_-\sigma _{\ast}}+\frac{i}{z}\sigma \eta e^{im_-\sigma _{\ast}}\sigma _3Q_{-}^{\ast}+\sum_{n=1}^{2N_1+2N_2+2N_3}{\frac{\underset{z=\widehat{\xi }_n}{\text{Res}}\,\,W^-}{z-\widehat{\xi }_n}}+\sum_{n=1}^{2N_1+2N_2+2N_3}{\frac{\underset{z=\xi _n}{\text{Res}}\,\,W^+}{z-\xi _n}}\\
				&+\sum_{n=2N_1+1}^{2N_1+2N_2+2N_3}{\frac{\underset{z=\hat{\xi}_n}{P_{-,2}}\,\,\,W^-}{\left( z-\widehat{\xi }_n \right) ^2}}+\sum_{n=2N_1+1}^{2N_1+2N_2+2N_3}{\frac{\underset{z=\xi _n}{P_{-,2}}\,\,W^+}{\left( z-\xi _n \right) ^2}}+\sum_{n=2N_1+2N_2+1}^{2N_1+2N_2+2N_3}{\frac{\underset{z=\hat{\xi}_n}{P_{-,3}}\,\,\,W^-}{\left( z-\widehat{\xi }_n \right) ^3}}\\
				&+\sum_{n=2N_1+2N_2+1}^{2N_1+2N_2+2N_3}{\frac{\underset{z=\xi _n}{P_{-,3}}\,\,W^+}{\left( z-\xi _n \right) ^3}}+\frac{1}{2\pi i}\int_{\varSigma}{\frac{W^+\left( s \right) J\left( s \right)}{s-z}}ds.\\
			\end{aligned}
		\end{equation}
Then, by comparing Eq.~(\ref{e15}) with the element from the first row and first column of Eq.~(\ref{e10}), we can derive the reconstruction formula for the reflectionless potentials, i.e.,
\begin{equation}
	\begin{aligned}
q&=q_-e^{2im_-}+ie^{im_-}\left( \sum_{k=1}^{_{N_1}}{A\left( \xi _k \right) \nu _{-12}\left( \xi _k \right)}+\sum_{k=N_1+1}^{_{N_1+N_2}}{\left( B\left( \xi _k \right) C\left( \xi _k \right) \nu _{-12}\left( \xi _k \right) +B\left( \xi _k \right) \nu _{-12}^{'}\left( \xi _k \right) \right)}\right.\\
&\left.+\sum_{k=N_1+N_2+1}^{_{N_1+N_2+N_3}}{\left( D\left( \xi _k \right) F\left( \xi _k \right) \nu _{-12}\left( \xi _k \right) +D\left( \xi _k \right) E\left( \xi _k \right) \nu _{-12}^{'}\left( \xi _k \right) +\frac{D\left( \xi _k \right)}{2}\nu _{-12}^{''}\left( \xi _k \right) \right)} \right). 
\end{aligned}
\end{equation}
Next, we need to solve for $\nu _{-12}\left( \xi _k \right)$, $\nu _{-12}^{'}\left( \xi _k \right)$, and $\nu _{-12}^{''}\left( \xi _k \right)$. By setting $W=W^-$ in Eq.~(\ref{e10}) and considering the element from the first row and first column, we have
 \begin{equation}
 	\begin{aligned}\label{e11}
 		\nu_{-11}\left( \widehat{\xi_k} \right) &= -\frac{iq_-e^{im_-}}{\widehat{\xi_k}}+ \sum_{n=1}^{N_1} \frac{A\left( \xi_n \right)\nu_{-12}\left( \xi_n \right)}{\widehat{\xi_k} - \xi_n} + \sum_{n=N_1+1}^{N_1+N_2} \left[\left( \frac{B\left( \xi_n \right)}{\left( \widehat{\xi_k} - \xi_n \right)^2} + \frac{B\left( \xi_n \right) C\left( \xi_n \right)}{\widehat{\xi_k} - \xi_n} \right) \nu_{-12}\left( \xi_n \right) \right. \\
 		& \left. + \frac{B\left( \xi_n \right)\nu_{-12}^{'}\left( \xi_{n} \right)}{\widehat{\xi_k} - \xi_n} \right]+\sum_{n=N_1+N_2+1}^{_{N_1+N_2+N_3}}\left[ \left( \frac{D\left( \xi _n \right) E\left( \xi _n \right)}{\widehat{\xi _k}-\xi _n}+\frac{D\left( \xi _n \right)}{\left( \widehat{\xi _k}-\xi _{n} \right) ^2} \right) \nu _{-12}^{'}\left( \xi _{n} \right)  \right.\\ 
 		&\left. +\frac{D\left( \xi _n \right)\nu _{-12}^{''}\left( \xi _n \right)}{2\left( \widehat{\xi _k}-\xi _n \right)}+\left( \frac{D\left( \xi _n \right) F\left( \xi _n \right)}{\widehat{\xi _k}-\xi _n}+\frac{D\left( \xi _n \right) E\left( \xi _n \right)}{\left( \widehat{\xi _k}-\xi _n \right) ^2}+\frac{D\left( \xi _n \right)}{\left( \widehat{\xi _k}-\xi _n \right) ^3} \right) \nu _{-12}\left( \xi _n \right) \right],
 	\end{aligned}
 \end{equation}
  \begin{equation}
 	\begin{aligned}\label{e12}
   \nu _{-11}^{'}\left( \widehat{\xi _k} \right) 
  &=\frac{iq_-e^{im_-}}{\widehat{\xi _k}^2}-\sum_{n=1}^{N_1}{\frac{A\left( \xi _n \right)\nu _{-12}\left( \xi _n \right)}{\left( \widehat{\xi _k}-\xi _n \right) ^2}} -\sum_{n=N_1+1}^{_{N_1+N_2}}\left[ \left( \frac{2B\left( \xi _n \right)}{\left( \widehat{\xi _k}-\xi _n \right) ^3}+\frac{B\left( \xi _n \right) C\left( \xi _n \right)}{\left( \widehat{\xi _k}-\xi _n \right) ^2} \right) \nu _{-12}\left( \xi _n \right) \right.\\
 	&\left.+\frac{B\left( \xi _n \right)\nu _{-12}^{'}\left( \xi _n \right)}{\left( \widehat{\xi _k}-\xi _n \right) ^2} \right]-\sum_{n=N_1+N_2+1}^{_{N_1+N_2+N_3}}\left[\left( \frac{D\left( \xi _n \right) E\left( \xi _n \right)}{\left( \widehat{\xi _k}-\xi _n \right) ^2}+\frac{2D\left( \xi _n \right)}{\left( \widehat{\xi _k}-\xi _n \right) ^3} \right) \nu _{-12}^{'}\left( \xi _n \right) \right.\\
 	&\left.  +\frac{D\left( \xi _n \right)\nu _{-12}^{''}\left( \xi _n \right)}{2\left( \widehat{\xi _k}-\xi _n \right) ^2}+ \left( \frac{D\left( \xi _n \right) F\left( \xi _n \right)}{\left( \widehat{\xi _k}-\xi _n \right) ^2}+\frac{2D\left( \xi _n \right) E\left( \xi _n \right)}{\left( \widehat{\xi _k}-\xi _n \right) ^3}+\frac{3D\left( \xi _n \right)}{\left( \widehat{\xi _k}-\xi _n \right) ^4} \right) \nu _{-12}\left( \xi _n \right) \right],
	\end{aligned}
\end{equation} 
   \begin{equation}
 	\begin{aligned}\label{e13}
 		\nu _{-11}^{''}\left( \widehat{\xi _k} \right)  
 		&=-\frac{2iq_-e^{im_-}}{\widehat{\xi _k}^3}+\sum_{n=1}^{N_1}{\frac{2A\left( \xi _n \right)\nu _{-12}\left( \xi _n \right)}{\left( \widehat{\xi _k}-\xi _n \right) ^3}} +\sum_{n=N_1+1}^{_{N_1+N_2}}\left[ \left( \frac{6B\left( \xi _n \right)}{\left( \widehat{\xi _k}-\xi _n \right) ^4}+\frac{2B\left( \xi _n \right) C\left( \xi _n \right)}{\left( \widehat{\xi _k}-\xi _n \right) ^3} \right) \nu _{-12}\left( \xi _n \right) \right.\\
 			&\left.+\frac{2B\left( \xi _n \right)\nu _{-12}^{'}\left( \xi _n \right)}{\left( \widehat{\xi _k}-\xi _n \right) ^3} \right] +\sum_{n=N_1+N_2+1}^{_{N_1+N_2+N_3}}\left[ \left( \frac{2D\left( \xi _n \right) E\left( \xi _n \right)}{\left( \widehat{\xi _k}-\xi _n \right) ^3}+\frac{6D\left( \xi _n \right)}{\left( \widehat{\xi _k}-\xi _n \right) ^4} \right) \nu _{-12}^{'}\left( \xi _n \right)\right.\\
 			&\left.  +\frac{D\left( \xi _n \right)\nu _{-12}^{''}\left( \xi _n \right)}{\left( \widehat{\xi _k}-\xi _n \right) ^3}+\left( \frac{2D\left( \xi _n \right) F\left( \xi _n \right)}{\left( \widehat{\xi _k}-\xi _n \right) ^3}+\frac{6D\left( \xi _n \right) E\left( \xi _n \right)}{\left( \widehat{\xi _k}-\xi _n \right) ^4}+\frac{12D\left( \xi _n \right)}{\left( \widehat{\xi _k}-\xi _n \right) ^5} \right) \nu _{-12}\left( \xi _n \right) \right].		
 	\end{aligned}
 \end{equation}
 Combining the symmetries of the Jost eigenfunctions, $\nu _{-11}\left( \widehat{\xi _k} \right)$, $\nu _{-11}^{'}\left( \widehat{\xi _k} \right)$, and $ \nu _{-11}^{''}\left( \widehat{\xi _k} \right)$ can be expressed by $\nu _{-12}\left( \xi _k \right)$, $\nu _{-12}^{'}\left( \xi _k \right)$, and $\nu _{-12}^{''}\left( \xi _k \right)$, i.e.,
\begin{align}
 &\nu _{-11}\left( \widehat{\xi _k} \right)=-\frac{iq_-}{\widehat{\xi _k}}\nu _{-12}\left( \xi _k \right) ,\\
 &\nu _{-11}^{'}\left( \widehat{\xi _k} \right) =\frac{iq_-}{\widehat{\xi _k}^2}\nu _{-12}\left( \xi _k \right) +\frac{i\sigma \eta q_-q_{0}^{2}}{\widehat{\xi _k}^3}\nu _{-12}^{'}\left( \xi _k \right),\\ \label{e14}
 &\nu _{-11}^{''}\left( \widehat{\xi _k} \right) =-\frac{2iq_-}{\widehat{\xi _k}^3}\nu _{-12}\left( \xi _k \right) -\frac{4i\sigma \eta q_-q_{0}^{2}}{\widehat{\xi _k}^4}\nu _{-12}^{'}\left( \xi _k \right) -\frac{iq_-q_{0}^{4}}{\widehat{\xi _k}^5}\nu _{-12}^{''}\left( \xi _k \right) .
\end{align}
 Then, we can obtain $\nu _{-12}\left( \xi _k \right)$, $\nu _{-12}^{'}\left( \xi _k \right)$, and $\nu _{-12}^{''}\left( \xi _k \right)$ from Eqs.~(\ref{e11})-(\ref{e14}), the expression for $q(x, t)$ can be written in matrix form:
	\begin{equation}
		\begin{aligned}
			q&=e^{2im_-}q_-(1+\frac{\det \left( \begin{matrix}
					0&		Y\\
					H&		G\\
				\end{matrix} \right)}{\det G}),
		\end{aligned}
	\end{equation}
	where\\
	\(Y_n=
	\begin{cases}
		A\left( \xi _n \right), & n\in \left[ 1,2N_1 \right],\\
		B\left( \xi _n \right) C\left( \xi _n \right), & n\in \left[2N_1+1,2N_1+2N_2 \right], \\
		B\left( \xi _n \right), & n\in \left[ 2N_1+2N_2+1,2N_1+4N_2 \right], \\
		D\left( \xi _n \right) F\left( \xi _n \right), & n\in \left[ 2N_1+4N_2+1,2N_1+4N_2+2N_3 \right], \\
		D\left( \xi _n \right) E\left( \xi _n \right), & n\in \left[ 2N_1+4N_2+2N_3+1,2N_1+4N_2+4N_3 \right], \\
		\dfrac{D\left( \xi _n \right)}{2}, & n\in \left[ 2N_1+4N_2+4N_3+1,2N_1+4N_2+6N_3 \right],
	\end{cases}\)\\
		\(H_k=
	\begin{cases}
		\frac{1}{\widehat{\xi _k}}, & k\in \left[ 1,2N_1+2N_2+2N_3 \right],\\
		\frac{1}{\widehat{\xi _k}^2}, & k\in \left[ 2N_1+2N_2+2N_3+1,2N_1+4N_2+4N_3 \right], \\
		\frac{2}{\widehat{\xi _k}^3}, & k\in \left[ 2N_1+4N_2+4N_3+1,2N_1+4N_2+6N_3 \right], 
	\end{cases}\)
\ $G=\left( \begin{array}{c}
		G_{k,n}^{\left( 1 \right)}\\
		G_{k,n}^{\left( 3 \right)}\\
		G_{k,n}^{\left( 5 \right)}\\
	\end{array}\begin{array}{c}
		G_{k,n}^{\left( 2 \right)}\\
		G_{k,n}^{\left( 4 \right)}\\
		G_{k,n}^{\left( 6 \right)}\\
	\end{array} \right) ,$

\begin{align*}
&G_{k,n}^{\left( 1 \right)}=\left\{ \begin{array}{l}
	\frac{A\left( \xi _n \right)}{\widehat{\xi _k}-\xi _n}+\frac{i\sigma \xi _k\delta _{k,n}}{q_{+}},\ k\in \left[ 1,2L \right] ,\ n\in \left[ 1,2N_1 \right] ,\\
	\frac{B\left( \xi _n \right)}{\left( \widehat{\xi _k}-\xi _n \right) ^2}+\frac{B\left( \xi _n \right) C\left( \xi _n \right)}{\widehat{\xi _k}-\xi _n}+\frac{i\sigma \xi _k\delta _{k,n}}{q_{+}},\ k\in \left[ 1,2L \right] ,\ n\in \left[ 2N_1+1,2N_1+2N_2 \right] ,\\
	\frac{B\left( \xi _n \right)}{\widehat{\xi _k}-\xi _n},\ k\in \left[ 1,2L \right] ,\ n\in \left[ 2N_1+2N_2+1,2N_1+4N_2 \right] ,\\
\end{array} \right. \\
&G_{k,n}^{\left( 2 \right)}=\left\{ \begin{array}{l}
	\frac{D\left( \xi _n \right) F\left( \xi _n \right)}{\widehat{\xi _k}-\xi _n}+\frac{D\left( \xi _n \right) E\left( \xi _n \right)}{\left( \widehat{\xi _k}-\xi _n \right) ^2}+\frac{D\left( \xi _n \right)}{\left( \widehat{\xi _k}-\xi _n \right) ^3}+\frac{i\sigma \xi _k\delta _{k,n}}{q_+},\ k\in \left[ 1,2L \right] ,\ n\in \left[ 2N_1+4N_2+1,2N_1+4N_2+2N_3 \right] ,\\
	\frac{D\left( \xi _n \right) E\left( \xi _n \right)}{\widehat{\xi _k}-\xi _n}+\frac{D\left( \xi _n \right)}{\left( \widehat{\xi _k}-\xi _n \right) ^2},\ k\in \left[ 1,2L \right] ,\ n\in \left[ 2N_1+4N_2+2N_3+1,2N_1+4N_2+4N_3 \right] ,\\
	\frac{D\left( \xi _n \right)}{2\left( \widehat{\xi _k}-\xi _n \right)},\ k\in \left[ 1,2L \right] ,\ n\in \left[ 2N_1+4N_2+4N_3+1,2N_1+4N_2+6N_3 \right] ,\\
\end{array} \right. 
\end{align*}
$$
G_{k,n}^{\left( 3 \right)}=\left\{ \begin{array}{l}
	\frac{A\left( \xi _n \right)}{\left( \widehat{\xi _k}-\xi _n \right) ^2},\ k\in \left[ 2L+1,2L+2N_2+2N_3 \right] ,\ n\in \left[ 1,2N_1 \right] ,\\
	\frac{2B\left( \xi _n \right)}{\left( \widehat{\xi _k}-\xi _n \right) ^3}+\frac{B\left( \xi _n \right) C\left( \xi _n \right)}{\left( \widehat{\xi _k}-\xi _n \right) ^2}+\frac{i\xi _k^2\delta _{k,n}}{q_{-}^{*}q_{0}^{2}},\ k\in \left[ 2L+1,2L+2N_2+2N_3 \right] ,\ n\in \left[ 2N_1+1,2N_1+2N_2 \right] ,\\
	\frac{B\left( \xi _n \right)}{\left( \widehat{\xi _k}-\xi _n \right) ^2}+\frac{i\xi _k^3\delta _{k,n}}{q_{-}^{*}q_{0}^{2}},\ k\in \left[ 2L+1,2L+2N_2+2N_3 \right] ,\ n\in \left[ 2N_1+2N_2+1,2N_1+4N_2 \right] ,\\
\end{array} \right. 
$$
$$
G_{k,n}^{\left( 4 \right)}=\left\{ \begin{array}{l}
	\frac{D\left( \xi _n \right) F\left( \xi _n \right)}{\left( \widehat{\xi _k}-\xi _n \right) ^2}+\frac{2D\left( \xi _n \right) E\left( \xi _n \right)}{\left( \widehat{\xi _k}-\xi _n \right) ^3}+\frac{3D\left( \xi _n \right)}{\left( \widehat{\xi _k}-\xi _n \right) ^4}+\frac{i\xi _k^2\delta _{k,n}}{q_{-}^{*}q_{0}^{2}},\ k\in \left[ 2L+1,2L+2N_2+2N_3 \right] ,\\
	\ \ \ \ \ \ \ \ \ \ \ \ \ \ \ \ \ \ \ \ \ \ \ \ \ \ \ \ \ \ \ \ \ \ \ \ \ \ \ \ \ \ \ \ \ \ \ \ \ \ \ \ \ \ \ \ n\in \left[ 2N_1+4N_2+1,2N_1+4N_2+2N_3 \right] ,\\
	\frac{D\left( \xi _n \right) E\left( \xi _n \right)}{\left( \widehat{\xi _k}-\xi _n \right) ^2}+\frac{2D\left( \xi _n \right)}{\left( \widehat{\xi _k}-\xi _n \right) ^3}+\frac{i\xi _k^3\delta _{k,n}}{q_{-}^{*}q_{0}^{2}},\ k\in \left[ 2L+1,2L+2N_2+2N_3 \right] ,\\
	\ \ \ \ \ \ \ \ \ \ \ \ \ \ \ \ \ \ \ \ \ \ \ \ \ \ \ \ \ \ \ \ \ \ \ \ \ \ \ n\in \left[ 2N_1+4N_2+2N_3+1,2N_1+4N_2+4N_3 \right] ,\\
	\frac{D\left( \xi _n \right)}{2\left( \widehat{\xi _k}-\xi _n \right) ^2},\ k\in \left[ 2L+1,2L+2N_2+2N_3 \right] ,\ n\in \left[ 2N_1+4N_2+4N_3+1,2N_1+4N_2+6N_3 \right] ,\\
\end{array} \right. 
$$
$$
G_{k,n}^{\left( 5 \right)}=\left\{ \begin{array}{l}
	\frac{2A\left( \xi _n \right)}{\left( \widehat{\xi _k}-\xi _n \right) ^3},\ k\in \left[ 2L+2N_2+2N_3+1,2L+2N_2+4N_3 \right] ,\ n\in \left[ 1,2N_1 \right] ,\\
	\frac{6B\left( \xi _n \right)}{\left( \widehat{\xi _k}-\xi _n \right) ^4}+\frac{2B\left( \xi _n \right) C\left( \xi _n \right)}{\left( \widehat{\xi _k}-\xi _n \right) ^3},\ k\in \left[ 2L+2N_2+2N_3+1,2L+2N_2+4N_3 \right] ,\ n\in \left[ 2N_1+1,2N_1+2N_2 \right] ,\\
	\frac{2B\left( \xi _n \right)}{\left( \widehat{\xi _k}-\xi _n \right) ^3},\ k\in \left[ 2L+2N_2+2N_3+1,2L+2N_2+4N_3 \right] ,\ n\in \left[ 2N_1+2N_2+1,2N_1+4N_2 \right] ,\\
\end{array} \right. 
$$
$$
G_{k,n}^{\left( 6 \right)}=\left\{ \begin{array}{l}
	\frac{2D\left( \xi _n \right) F\left( \xi _n \right)}{\left( \widehat{\xi _k}-\xi _n \right) ^3}+\frac{6D\left( \xi _n \right) E\left( \xi _n \right)}{\left( \widehat{\xi _k}-\xi _n \right) ^4}+\frac{12D\left( \xi _n \right)}{\left( \widehat{\xi _k}-\xi _n \right) ^5}+\frac{2i\sigma \xi _k^3\delta _{k,n}}{q_{+}q_{0}^{4}},\ k\in \left[ 2L+2N_2+2N_3+1,2L+2N_2+4N_3 \right] ,\\
	\ \ \ \ \ \ \ \ \ \ \ \ \ \ \ \ \ \ \ \ \ \ \ \ \ \ \ \ \ \ \ \ \ \ \ \ \ \ \ \ \ \ \ \ \ \ \ \ \ \ \ \ \ \ \ \ \ \ \ \ n\in \left[ 2N_1+4N_2+1,2N_1+4N_2+2N_3 \right] ,\\
	\frac{2D\left( \xi _n \right) E\left( \xi _n \right)}{\left( \widehat{\xi _k}-\xi _n \right) ^3}+\frac{6D\left( \xi _n \right)}{\left( \widehat{\xi _k}-\xi _n \right) ^4}+\frac{4i\sigma \xi _k^4\delta _{k,n}}{q_{+}q_{0}^{4}},\ k\in \left[ 2L+2N_2+2N_3+1,2L+2N_2+4N_3 \right] ,\\
	\ \ \ \ \ \ \ \ \ \ \ \ \ \ \ \ \ \ \ \ \ \ \ \ \ \ \ \ \ \ \ \ \ \ \ \ \ \ \ \ \ \ \ n\in \left[ 2N_1+4N_2+2N_3+1,2N_1+4N_2+4N_3 \right] ,\\
	\frac{D\left( \xi _n \right)}{\left( \widehat{\xi _k}-\xi _n \right) ^3}+\frac{i\sigma \xi _k^5\delta _{k,n}}{q_{+}q_{0}^{4}},\ k\in \left[ 2L+2N_2+2N_3+1,2L+2N_2+4N_3 \right] ,\\
	\ \ \ \ \ \ \ \ \ \ \ \ \ \ \ \ \ \ \ \ \ \ \ \ n\in \left[ 2N_1+4N_2+4N_3+1,2N_1+4N_2+6N_3 \right] .\\
\end{array} \right. 
$$
	Since $\sigma$ can be unified through a transformation, we will exclusively focus on the case that $\sigma=-1$. 
	
	Case 1: $\sigma=-1,\ \eta=1$. 
	In this case, $\theta$ condition (\ref{e5}) reduces to 
	\begin{equation}
		\frac{q_-^{\ast}}{q_-}=\prod_{n=1}^{N_1}{\frac{\xi _n^2}{\widehat{\xi _n}^2}}\prod_{n=N_1+1}^{N_1+N_2}{\frac{\xi _n^4}{\widehat{\xi _n}^4}}\prod_{n=N_1+N_2+1}^{N_1+N_2+N_3}{\frac{\xi _n^6}{\widehat{\xi _n}^6}}e^{2i\bar{m}}.
	\end{equation}

	$\bullet$ For three single poles, i.e., $N_1=3,\ N_2=0,\ N_3=0$, we set $q_-=1,\ \xi_{1,1}=e^{\frac{5i\pi}{24}},\ \xi_{1,2}=e^{\frac{9i\pi}{24}},\ \xi_{1,3}=e^{\frac{10i\pi}{24}}$, which represents the behavior of three-soliton as shown in Fig. 1. For four single poles, i.e., $N_1=4,\ N_2=0,\ N_3=0$, we consider $q_-=1,\ \xi_{1,1}=2e^{\frac{i\pi}{8}},\ \xi_{1,2}=\frac{1}{2}e^{\frac{i\pi}{8}},\ \xi_{1,3}=e^{\frac{i\pi}{3}},\ \xi_{1,4}=e^{\frac{5i\pi}{12}}$, which corresponds to the dynamics of a breather with a bright-bright soliton depicted in Fig. 2.
		\begin{figure}[H]
		\centering
		\begin{minipage}[c]{0.23\textwidth} 
			\centering
			\includegraphics[width=\textwidth]{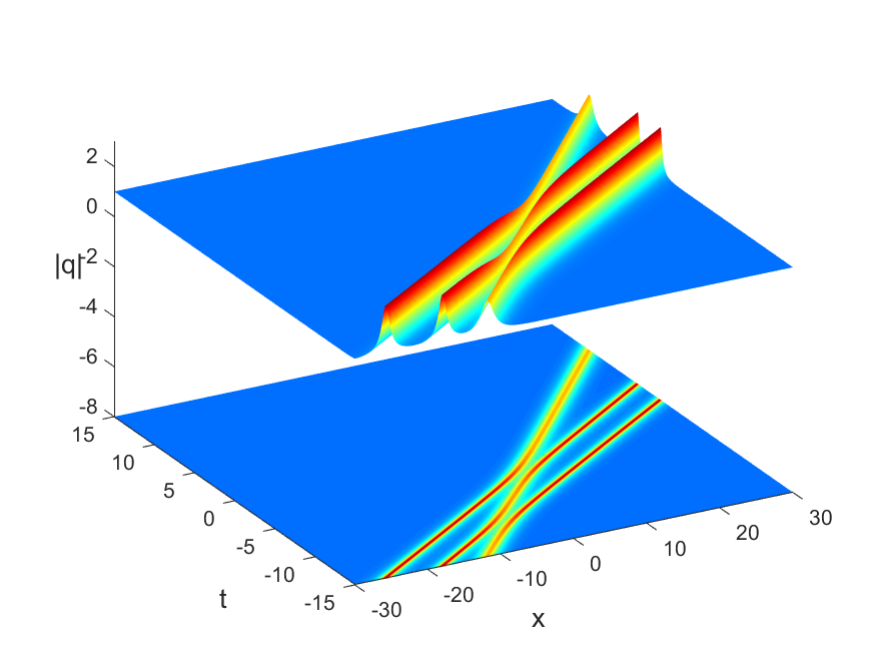} 
			\centerline{(a)}
		\end{minipage}
		\begin{minipage}[c]{0.23\textwidth}
			\centering
			\includegraphics[width=\textwidth]{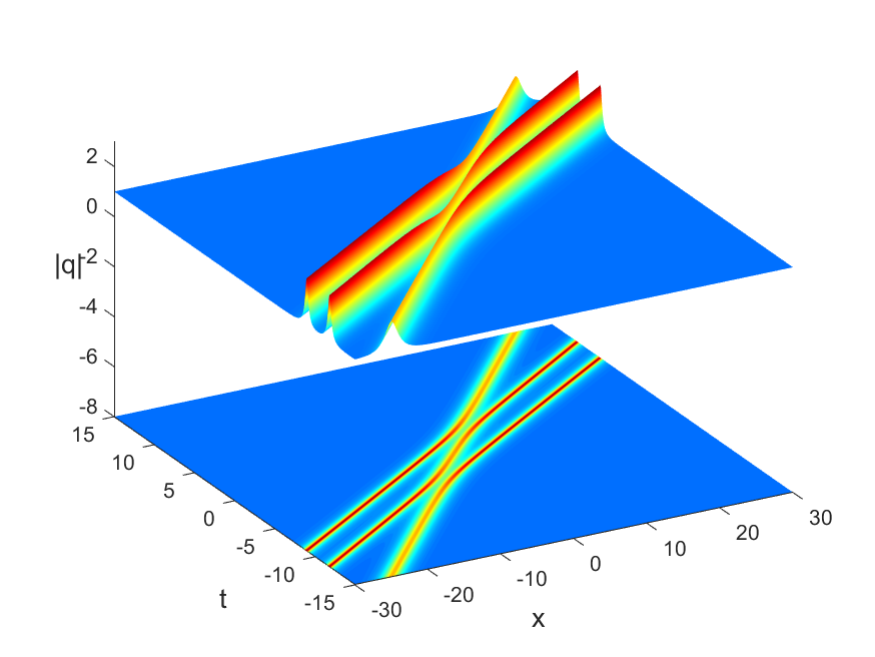}
			\centerline{(b)}
		\end{minipage}
		\begin{minipage}[c]{0.23\textwidth}
			\centering
			\includegraphics[width=\textwidth]{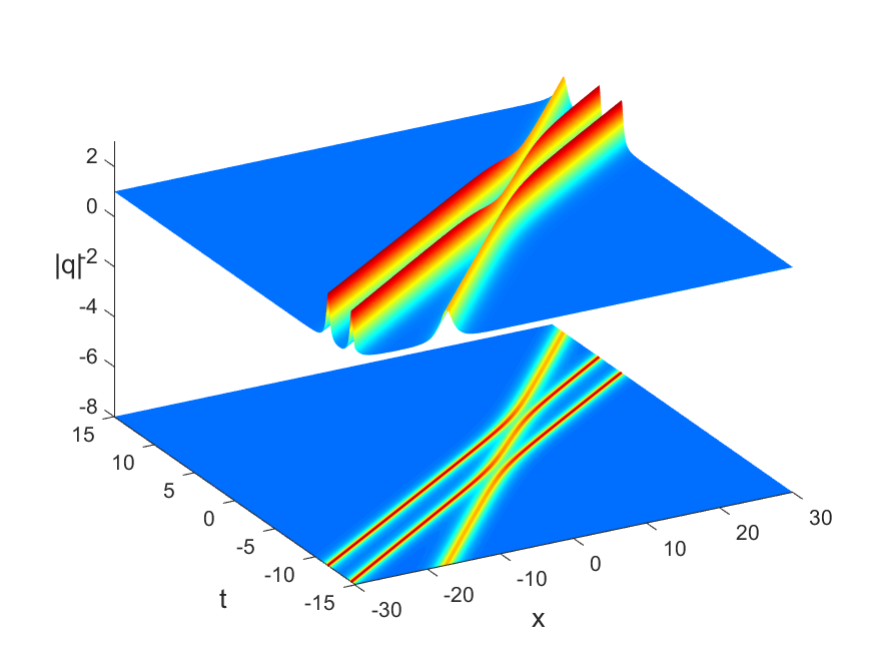}
			\centerline{(c)}
		\end{minipage}
		\begin{minipage}[c]{0.23\textwidth}
			\centering
			\includegraphics[width=\textwidth]{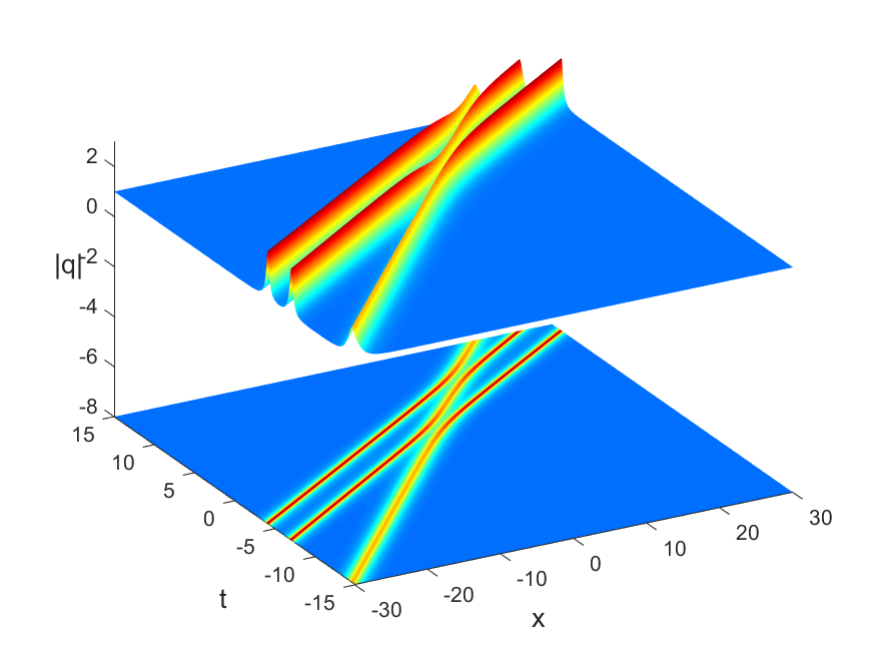}
			\centerline{(d)}
		\end{minipage}
		\caption{Three-soliton with $q_-=1,\ \xi_{1,1}=e^{\frac{5i\pi}{24}},\ \xi_{1,2}=e^{\frac{9i\pi}{24}},\ \xi_{1,3}=e^{\frac{10i\pi}{24}}$. (a) $x_0=0,\ t_0=-15$, (b) $x_0=0,\ t_0=0$, (c) $x_0=15,\ t_0=0$, (d) $x_0=15,\ t_0=15$.}
	\end{figure}
		\begin{figure}[H]
		\centering
		\begin{minipage}[c]{0.23\textwidth} 
			\centering
			\includegraphics[width=\textwidth]{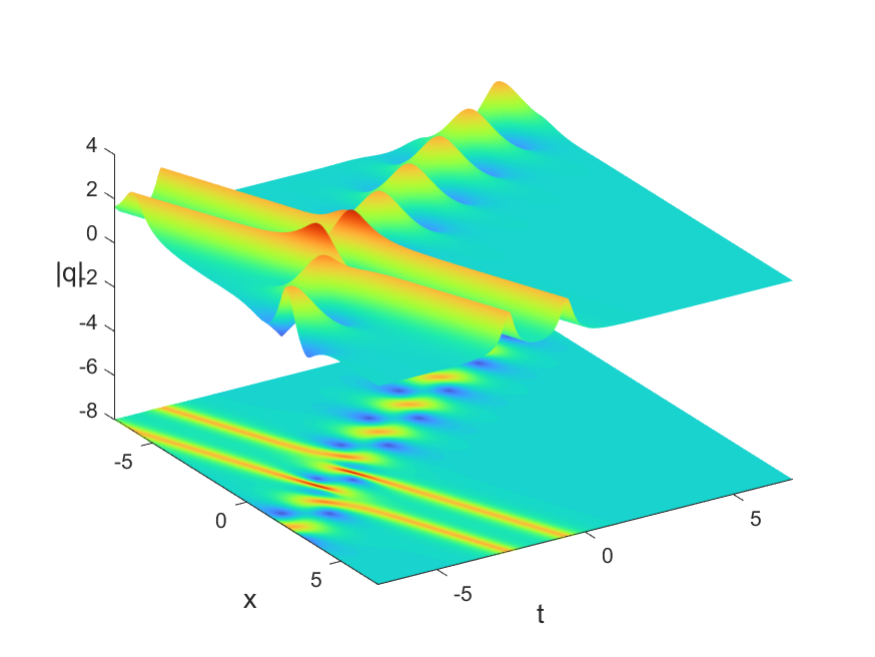} 
			\centerline{(a)}
		\end{minipage}
		\begin{minipage}[c]{0.23\textwidth}
			\centering
			\includegraphics[width=\textwidth]{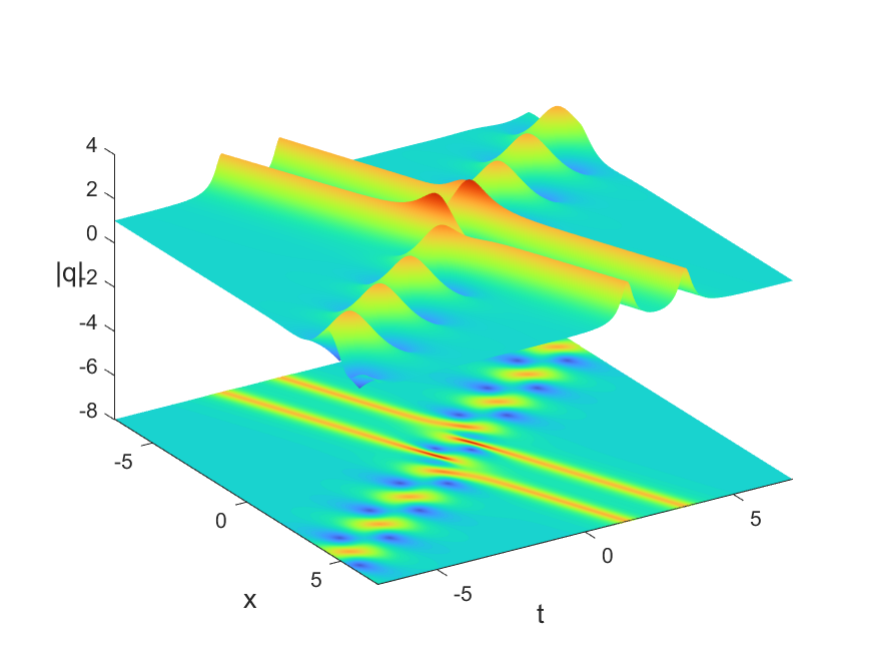}
			\centerline{(b)}
		\end{minipage}
		\begin{minipage}[c]{0.23\textwidth}
			\centering
			\includegraphics[width=\textwidth]{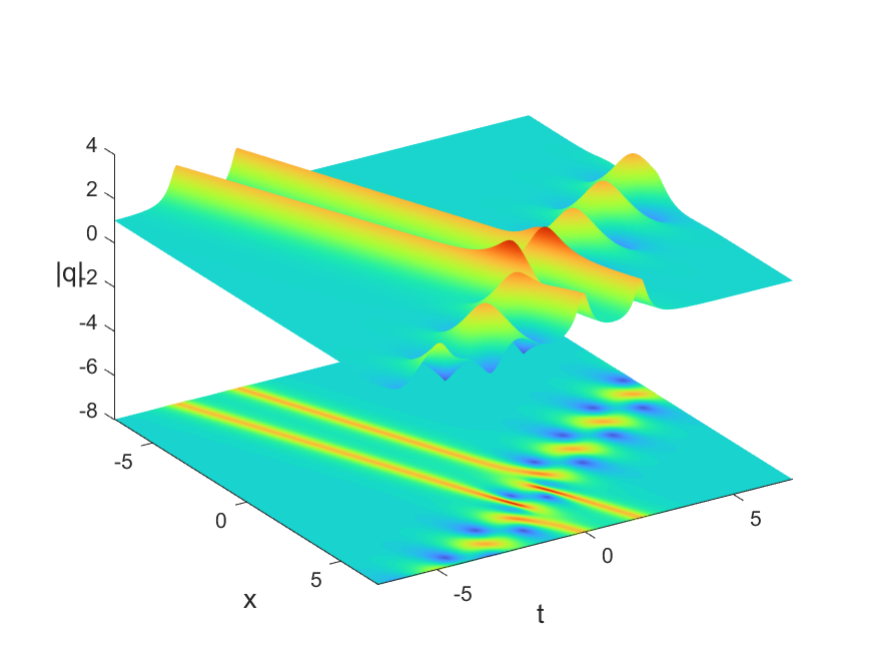}
			\centerline{(c)}
		\end{minipage}
		\begin{minipage}[c]{0.23\textwidth}
			\centering
			\includegraphics[width=\textwidth]{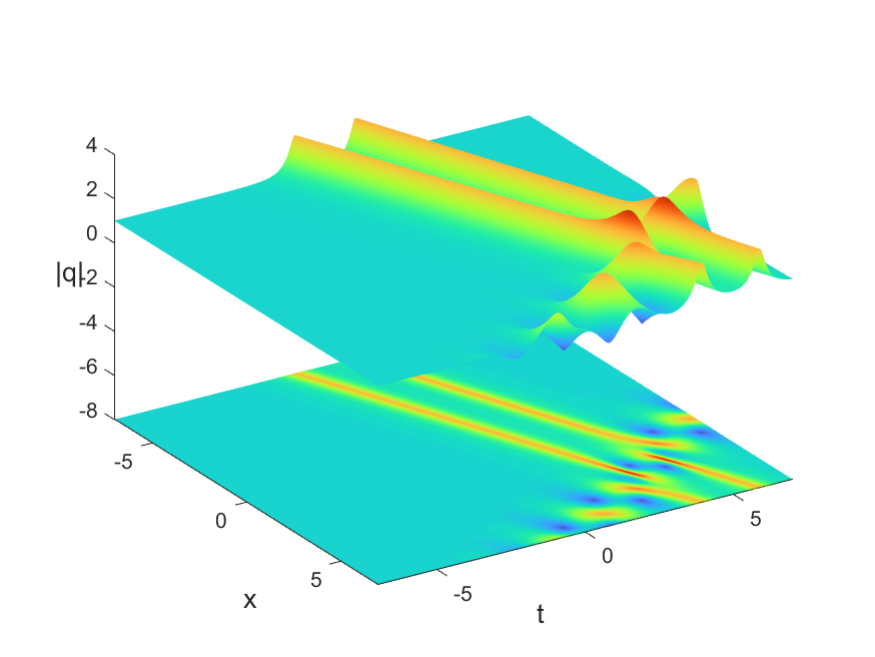}
			\centerline{(d)}
		\end{minipage}
		\caption{An interaction between a breather and a bright-bright soliton with $q_-=1,\ \xi_{1,1}=2e^{\frac{i\pi}{8}},\ \xi_{1,2}=\frac{1}{2}e^{\frac{i\pi}{8}},\ \xi_{1,3}=e^{\frac{i\pi}{3}},\ \xi_{1,4}=e^{\frac{5i\pi}{12}}$. (a) $x_0=0,\ t_0=-8$, (b) $x_0=0,\ t_0=0$, (c) $x_0=8,\ t_0=0$, (d) $x_0=8,\ t_0=8$.}
	\end{figure}
	$\bullet$ For two single poles and a double pole, i.e., $N_1=2,\ N_2=1,\ N_3=0$, we consider $q_-=1,\ \xi_{1,1}=2e^{\frac{i\pi}{8}},\ \xi_{1,2}=\frac{1}{2}e^{\frac{i\pi}{8}},\ \xi_{2,1}=e^{\frac{i\pi}{8}}$, which represents the dynamics of a breather with a bright-dark soliton illustrated in Fig. 3.
	\begin{figure}[H]
		\centering
		\begin{minipage}[c]{0.23\textwidth} 
			\centering
			\includegraphics[width=\textwidth]{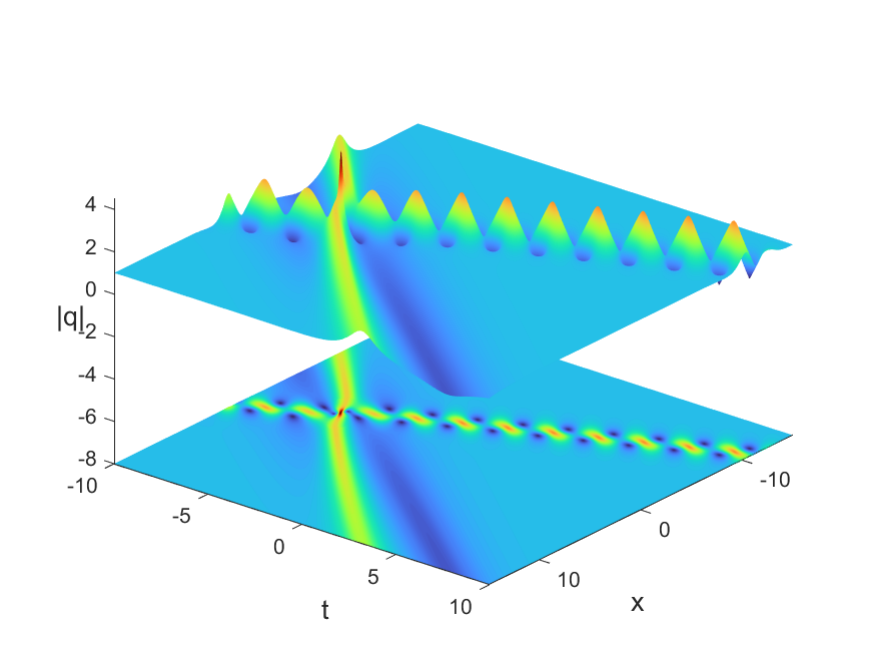} 
			\centerline{(a)}
		\end{minipage}
		\begin{minipage}[c]{0.23\textwidth}
			\centering
			\includegraphics[width=\textwidth]{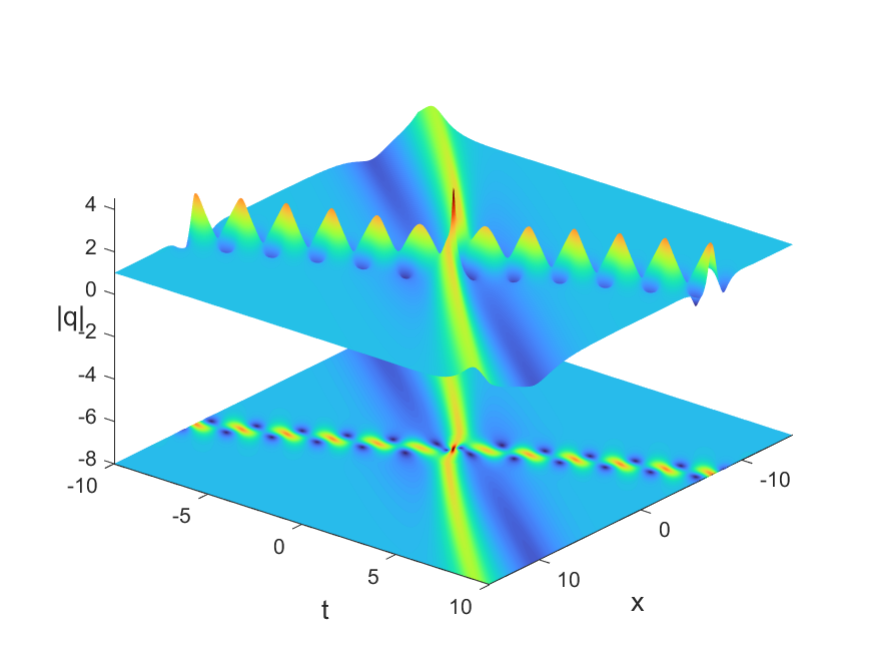}
			\centerline{(b)}
		\end{minipage}
		\begin{minipage}[c]{0.23\textwidth}
			\centering
			\includegraphics[width=\textwidth]{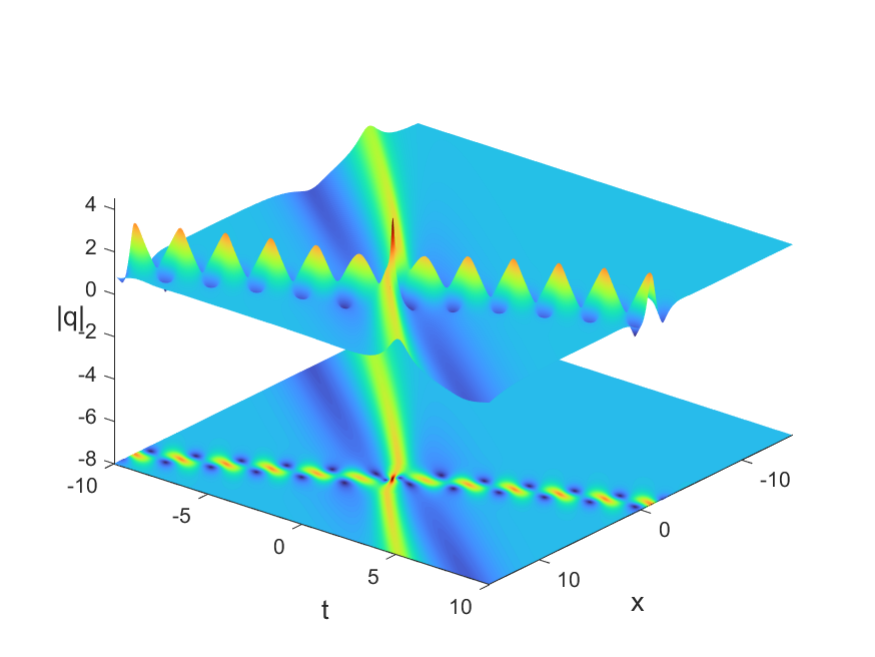}
			\centerline{(c)}
		\end{minipage}
		\begin{minipage}[c]{0.23\textwidth}
			\centering
			\includegraphics[width=\textwidth]{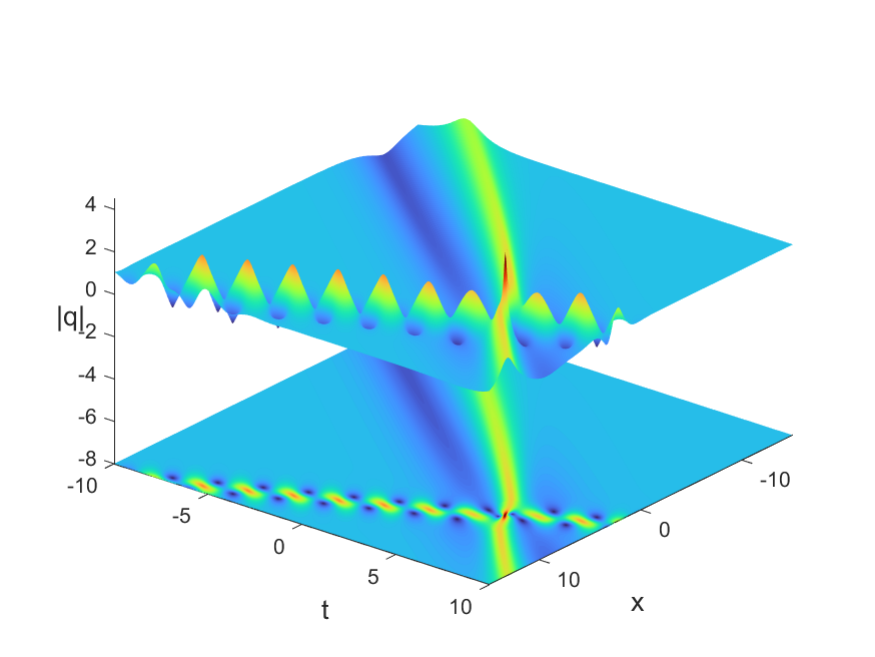}
			\centerline{(d)}
		\end{minipage}
		\caption{An interaction between a breather and a bright-dark soliton with $q_-=1,\ \xi_{1,1}=2e^{\frac{i\pi}{8}},\ \xi_{1,2}=\frac{1}{2}e^{\frac{i\pi}{8}},\ \xi_{2,1}=e^{\frac{i\pi}{8}}$. (a) $x_0=0,\ t_0=-10$, (b) $x_0=0,\ t_0=0$, (c) $x_0=10,\ t_0=0$, (d) $x_0=10,\ t_0=10$.}
	\end{figure}
		$\bullet$ For two double poles, i.e., $N_1=0,\ N_2=2,\ N_3=0$, we choose $q_-=1,\ \xi_{2,1}=2e^{\frac{i\pi}{8}},\ \xi_{2,2}=\frac{1}{2}e^{\frac{i\pi}{8}}$, which relates to the dynamics of two-breather as shown in Fig. 4.
	\begin{figure}[H]
		\centering
		\begin{minipage}[c]{0.23\textwidth} 
			\centering
			\includegraphics[width=\textwidth]{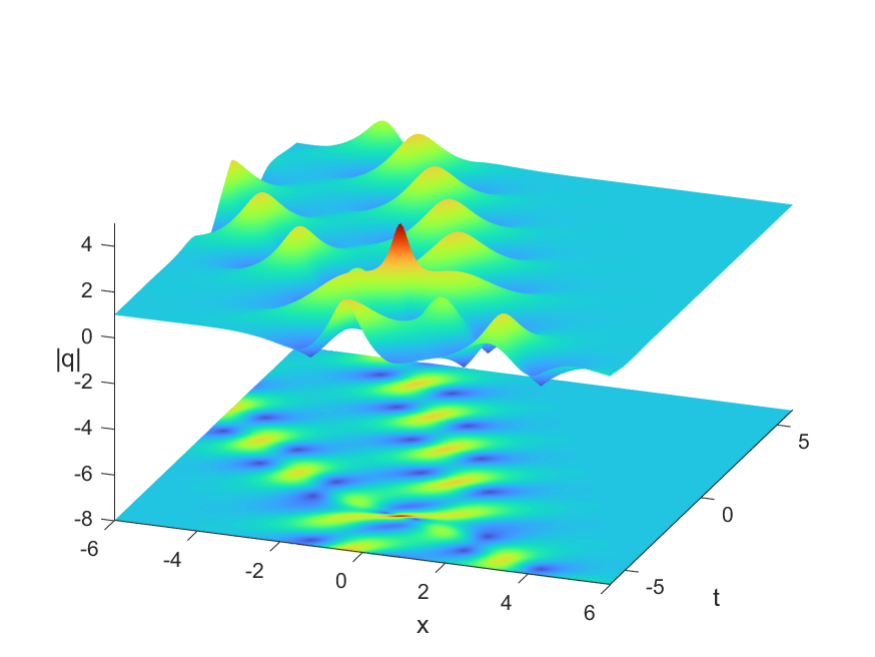} 
			\centerline{(a)}
		\end{minipage}
		\begin{minipage}[c]{0.23\textwidth}
			\centering
			\includegraphics[width=\textwidth]{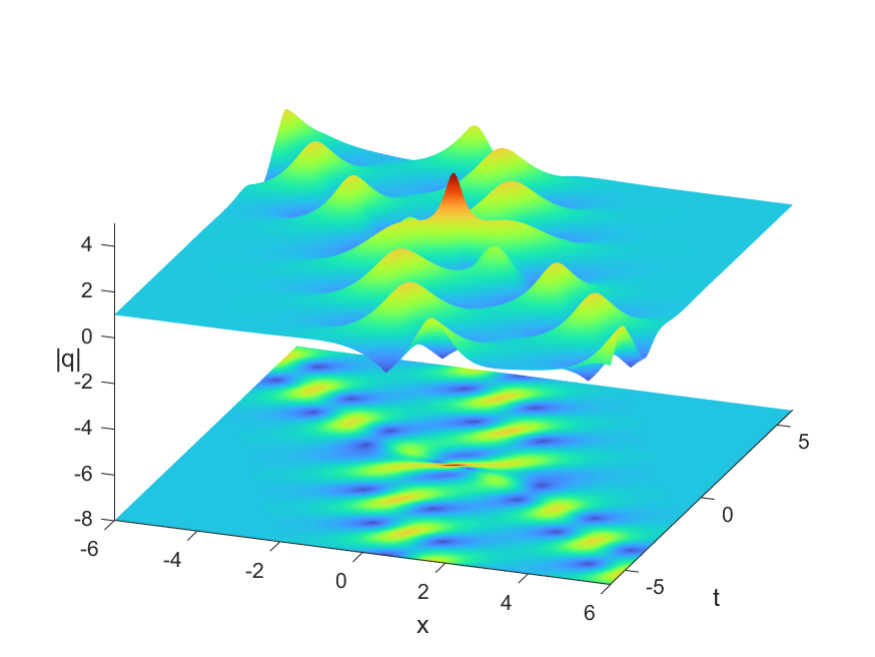}
			\centerline{(b)}
		\end{minipage}
		\begin{minipage}[c]{0.23\textwidth}
			\centering
			\includegraphics[width=\textwidth]{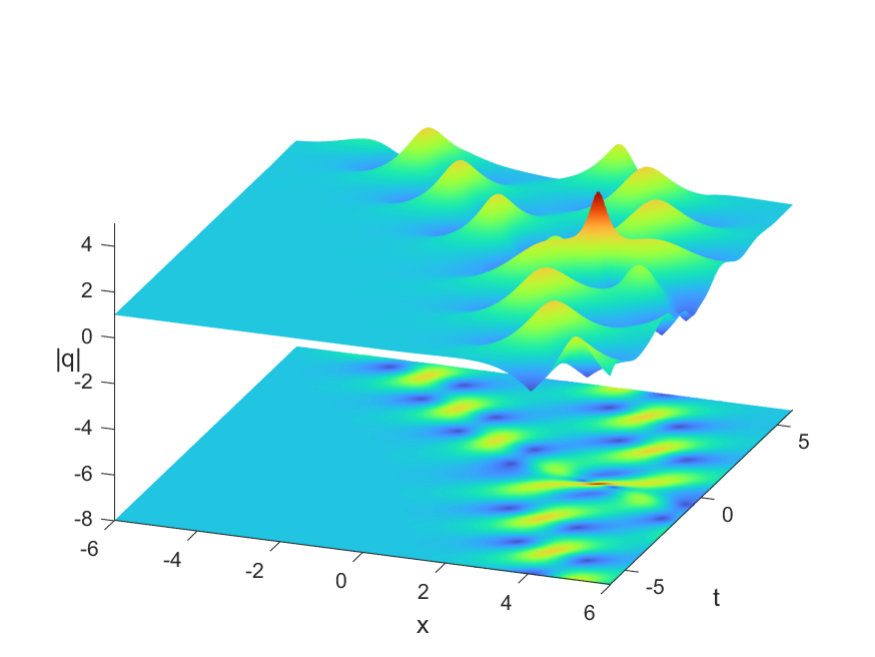}
			\centerline{(c)}
		\end{minipage}
		\begin{minipage}[c]{0.23\textwidth}
			\centering
			\includegraphics[width=\textwidth]{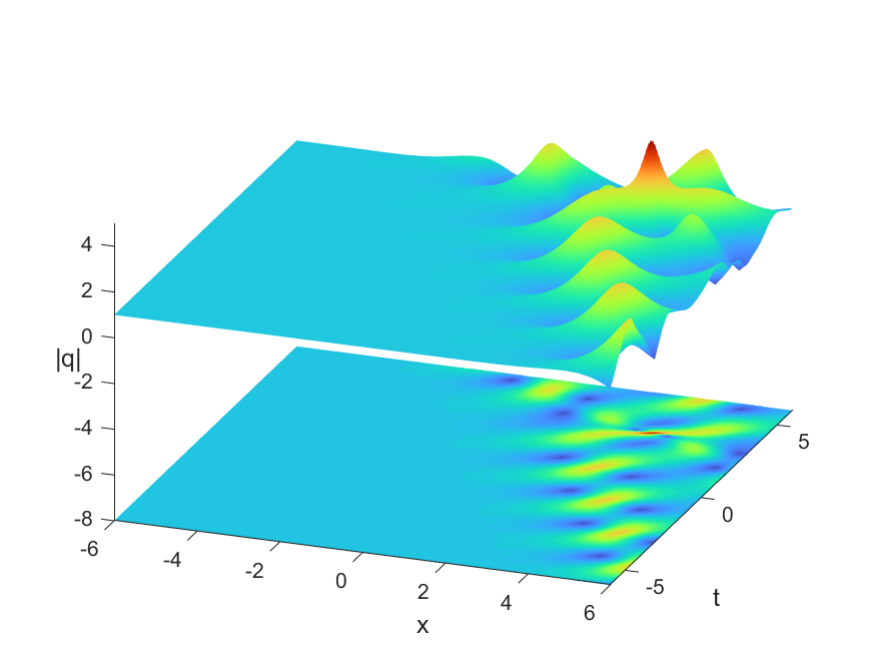}
			\centerline{(d)}
		\end{minipage}
		\caption{Two-breather with $q_-=1,\ \xi_{2,1}=2e^{\frac{i\pi}{8}},\ \xi_{2,2}=\frac{1}{2}e^{\frac{i\pi}{8}}$. (a) $x_0=0,\ t_0=-7$, (b) $x_0=0,\ t_0=0$, (c) $x_0=7,\ t_0=0$, (d) $x_0=7,\ t_0=7$.}
		\end{figure}
			$\bullet$ For two triple poles, i.e., $N_1=0,\ N_2=0,\ N_3=2$, we choose $q_-=1,\ \xi_{3,1}=2e^{\frac{i\pi}{6}},\ \xi_{3,2}=\frac{1}{2}e^{\frac{i\pi}{6}}$, which relates to the dynamics of three-breather as shown in Fig. 5.
		\begin{figure}[H]
			\centering
			\begin{minipage}[c]{0.23\textwidth} 
				\centering
				\includegraphics[width=\textwidth]{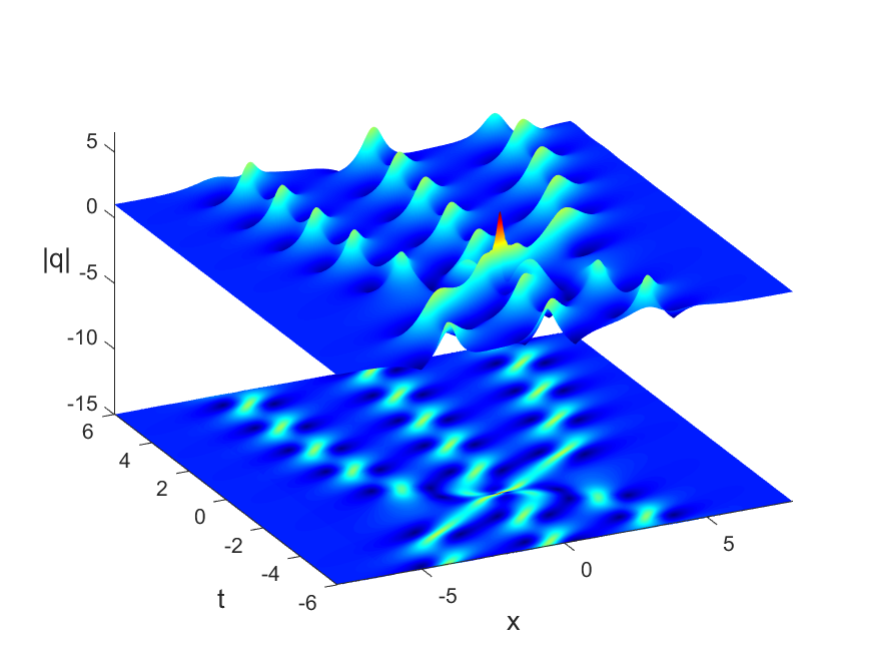} 
				\centerline{(a)}
			\end{minipage}
			\begin{minipage}[c]{0.23\textwidth}
				\centering
				\includegraphics[width=\textwidth]{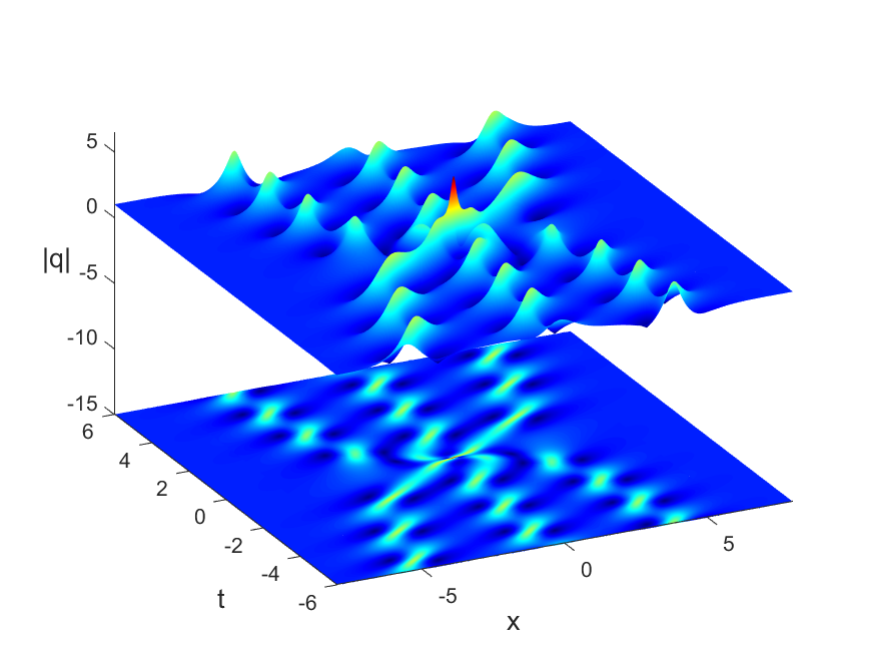}
				\centerline{(b)}
			\end{minipage}
			\begin{minipage}[c]{0.23\textwidth}
				\centering
				\includegraphics[width=\textwidth]{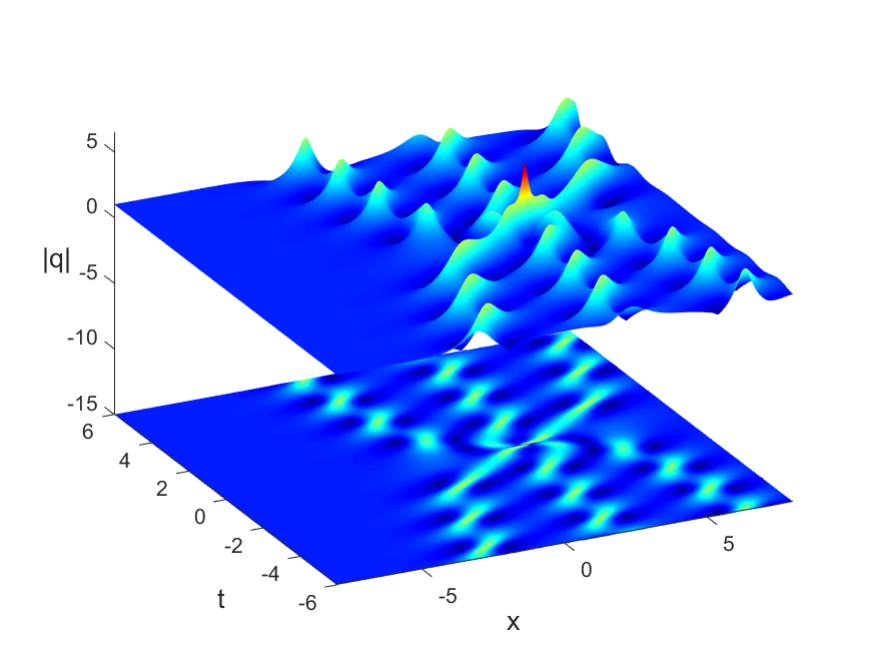}
				\centerline{(c)}
			\end{minipage}
			\begin{minipage}[c]{0.23\textwidth}
				\centering
				\includegraphics[width=\textwidth]{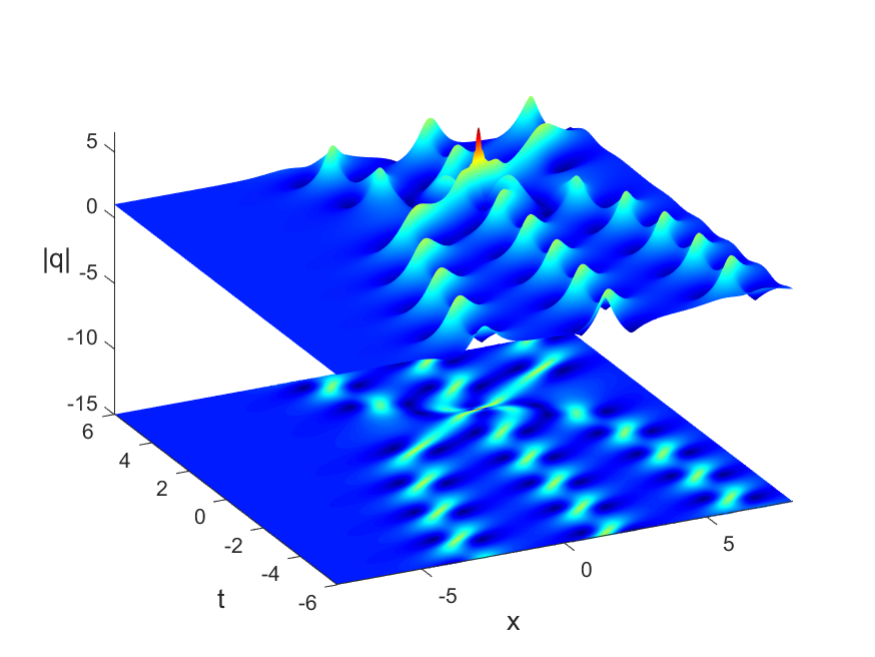}
				\centerline{(d)}
			\end{minipage}
			\caption{Three-breather with $q_-=1,\ \xi_{3,1}=2e^{\frac{i\pi}{6}},\ \xi_{3,2}=\frac{1}{2}e^{\frac{i\pi}{6}}$. (a) $x_0=0,\ t_0=-5$, (b) $x_0=0,\ t_0=0$, (c) $x_0=5,\ t_0=0$, (d) $x_0=5,\ t_0=5$.}
		\end{figure}
		Case 2: $\sigma=-1,\ \eta=-1$. 
In this case, $\theta$ condition (\ref{e5}) reduces to 
\begin{equation}
	-\frac{q_-^{\ast}}{q_-}=\prod_{n=1}^{N_1}{\frac{\xi _n^2}{\widehat{\xi _n}^2}}\prod_{n=N_1+1}^{N_1+N_2}{\frac{\xi _n^4}{\widehat{\xi _n}^4}}\prod_{n=N_1+N_2+1}^{N_1+N_2+N_3}{\frac{\xi _n^6}{\widehat{\xi _n}^6}}e^{2i\bar{m}}.
\end{equation}

$\bullet$ For three single poles, i.e., $N_1=3,\ N_2=0,\ N_3=0,$ we take $q_-=1,\ \xi_{1,1}=2e^{\frac{i\pi}{3}},\ \xi_{1,2}=\frac{1}{2}e^{\frac{i\pi}{3}},\ \xi_{1,3}=e^{\frac{i\pi}{12}}$, which corresponds to the interaction of a breather with a bright soliton presented in Fig. 6.
\begin{figure}[H]
	\centering
	\begin{minipage}[c]{0.23\textwidth} 
		\centering
		\includegraphics[width=\textwidth]{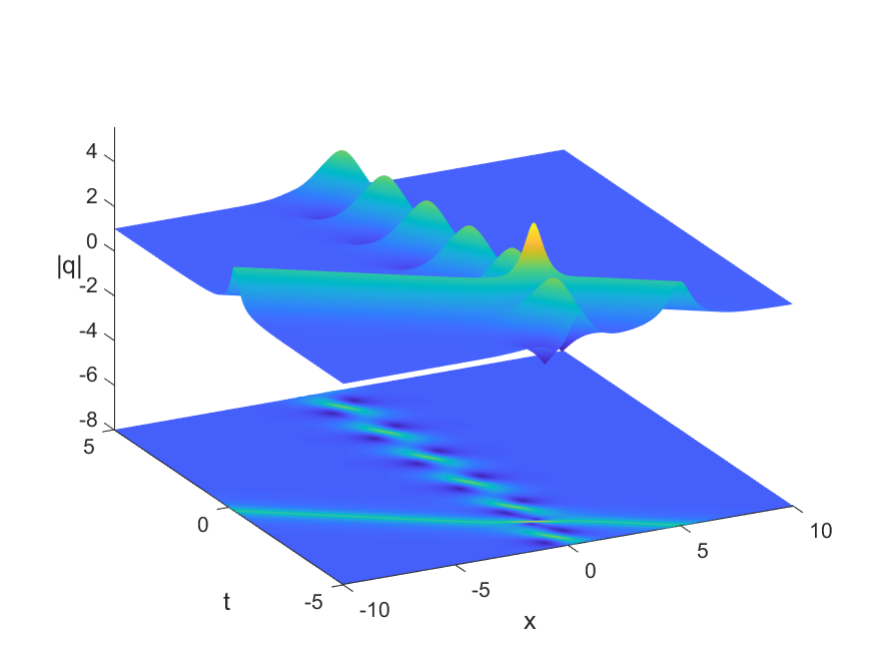} 
		\centerline{(a)}
	\end{minipage}
	\begin{minipage}[c]{0.23\textwidth}
		\centering
		\includegraphics[width=\textwidth]{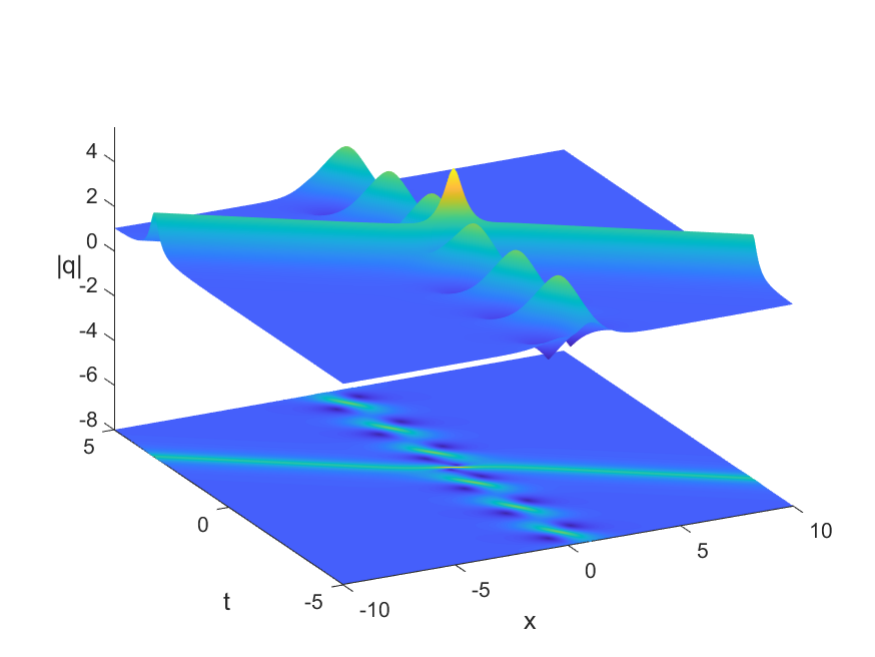}
		\centerline{(b)}
	\end{minipage}
	\begin{minipage}[c]{0.23\textwidth}
		\centering
		\includegraphics[width=\textwidth]{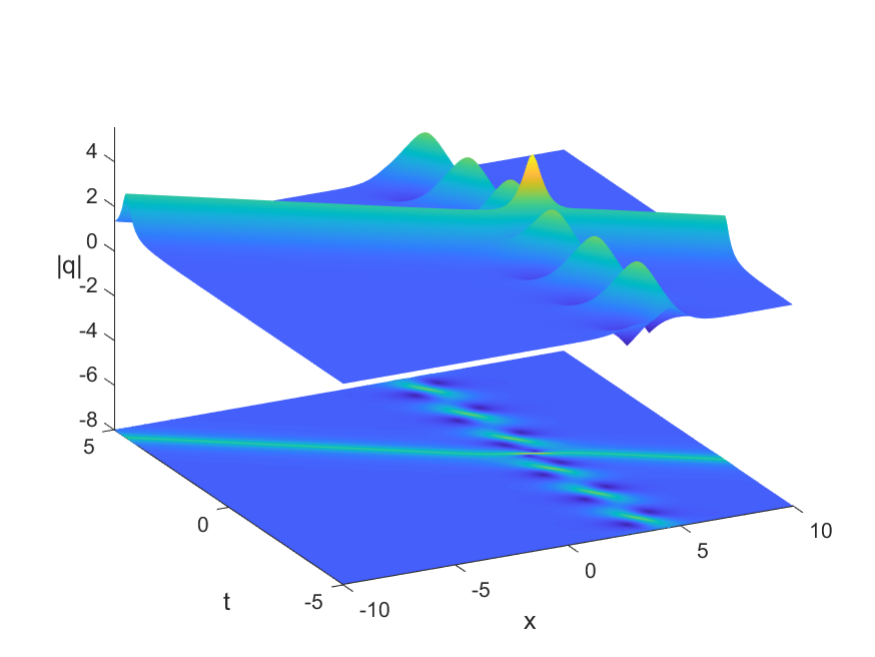}
		\centerline{(c)}
	\end{minipage}
	\begin{minipage}[c]{0.23\textwidth}
		\centering
		\includegraphics[width=\textwidth]{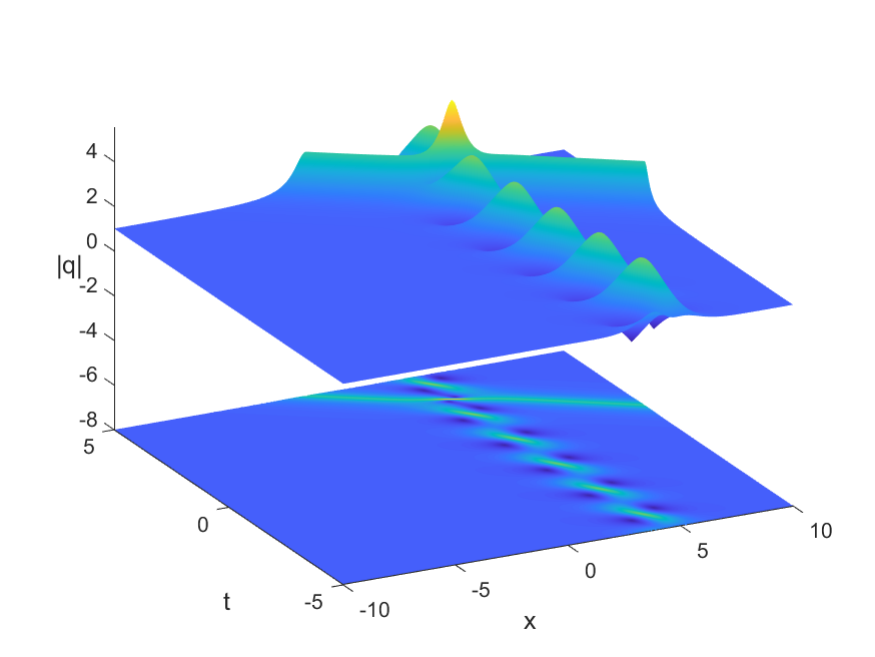}
		\centerline{(d)}
	\end{minipage}
	\caption{An interaction between a breather and a bright soliton with $q_-=1,\ \xi_{1,1}=2e^{\frac{i\pi}{3}},\ \xi_{1,2}=\frac{1}{2}e^{\frac{i\pi}{3}},\ \xi_{1,3}=e^{\frac{i\pi}{12}}$. (a) $x_0=0,\ t_0=-7$, (b) $x_0=0,\ t_0=0$, (c) $x_0=7,\ t_0=0$, (d) $x_0=7,\ t_0=7$.}
\end{figure}
	$\bullet$ For a triple pole, i.e., $N_1=0,\ N_2=0,\ N_3=1$, we assume $q_-=1,\ \xi_{3,1}=e^{\frac{i\pi}{4}}$, which corresponds to the behavior of a birght-dark-bright soliton as illustrated in Fig. 7.
		\begin{figure}[H]
			\centering
			\begin{minipage}[c]{0.23\textwidth} 
				\centering
				\includegraphics[width=\textwidth]{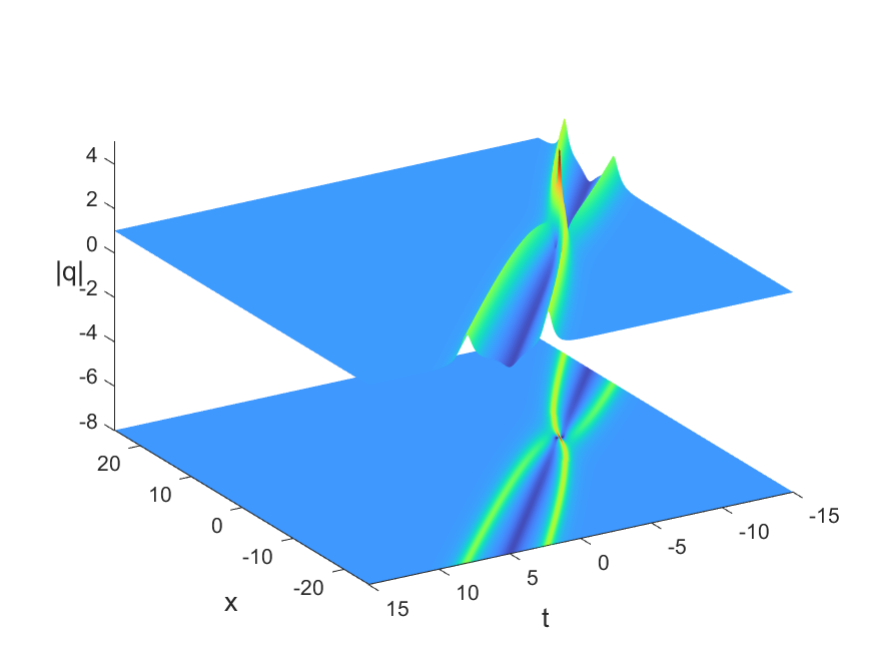} 
				\centerline{(a)}
			\end{minipage}
			\begin{minipage}[c]{0.23\textwidth}
				\centering
				\includegraphics[width=\textwidth]{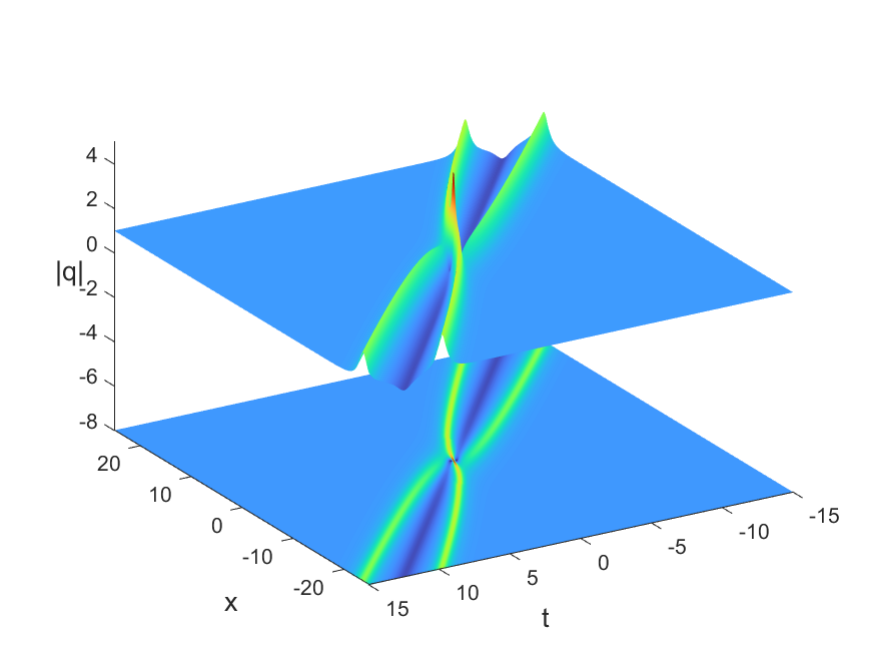}
				\centerline{(b)}
			\end{minipage}
			\begin{minipage}[c]{0.23\textwidth}
				\centering
				\includegraphics[width=\textwidth]{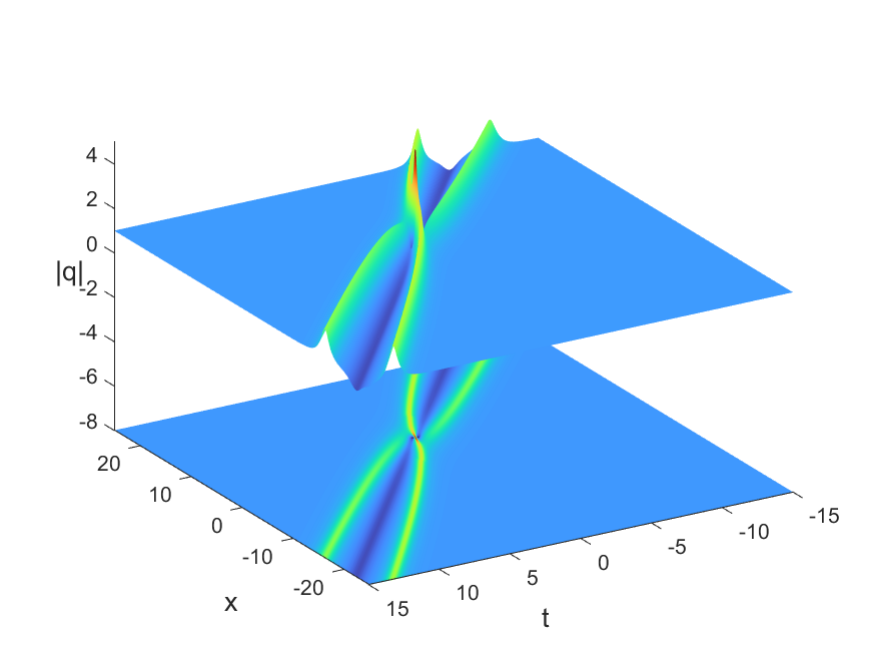}
				\centerline{(c)}
			\end{minipage}
			\begin{minipage}[c]{0.23\textwidth}
				\centering
				\includegraphics[width=\textwidth]{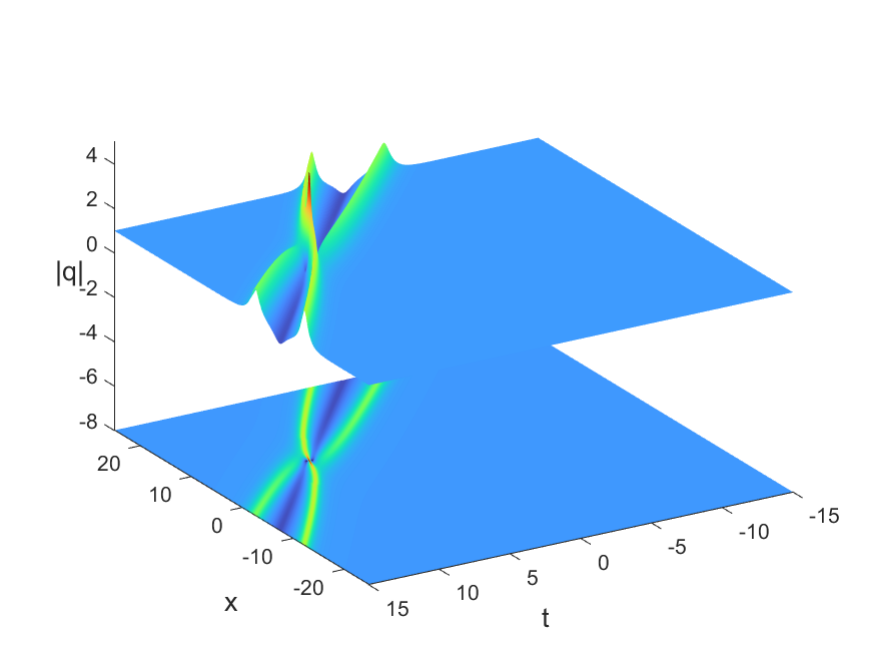}
				\centerline{(d)}
			\end{minipage}
			\caption{A bright-dark-bright soliton with $q_-=1,\ \xi_{3,1}=e^{\frac{i\pi}{4}}$. (a) $x_0=0,\ t_0=-15$, (b) $x_0=0,\ t_0=0$, (c) $x_0=15,\ t_0=0$, (d) $x_0=15,\ t_0=15$.}
		\end{figure}		
		Through the observation and analysis of the graphs above, it can be seen that different types of solutions depend on the choice of discrete eigenvalues. When we select a discrete eigenvalue on a quarter circle of radius 1 in the first quadrant of the complex $z$ plane, the discrete eigenvalue is a single pole, a double pole, or a triple pole, corresponding to one-soliton, two-soliton, and three-soliton, respectively. When we select two discrete eigenvalues in the first quadrant of the complex $z$ plane, one discrete eigenvalue is located within a quarter circle of radius 1, and the other discrete eigenvalue is located outside the quarter circle, the discrete eigenvalues are two single poles, two double poles, or two triple poles, corresponding to one-breather, two-breather, and three-breather, respectively. Additionally, we can also observe that the shift parameters $x_0$ and $t_0$ don't change the amplitude and propagation direction of solitons, but only change the position of the interaction between solitons.
	\section{Conclusions}
	\hspace{0.7cm}In this paper, we have investigated the mixed single, double, and triple poles solutions of the space-time shifted nonlocal DNLS equation under NZBCs via the RH approach. First, the space-time shifted nonlocal DNLS equation (\ref{e1}) has been derived by imposing the space-time shifted nonlocal integrable symmetry reduction $r(x,t)=\sigma q(x_{0}-x,t_{0}-t)$. Then, the analyticity, symmetries, and asymptotic behaviors of the Jost eigenfunctions and scattering matrix functions have been analyzed. Furthermore, by solving the RH problem, the expression for the solution with the reflectionless potentials has been given for the case where the single, double, and triple poles exist simultaneously. Finally, we have employed graphical simulation to illustrate the effects of shift parameters $x_{0}, t_{0}$ and present several representative solutions, including three-soliton, two-breather, and soliton-breather solutions.
	
	\vspace{5mm}\noindent\textbf{Data availability}
	\\\hspace*{\parindent}
	
	No data was used for the research described in the article.
	
	\vspace{5mm}\noindent\textbf{Acknowledgements}
	\\\hspace*{\parindent}We express our sincere thanks to each
	member of our discussion group for their suggestions. This work has been supported by the Fund Program for the Scientific Activities of Selected Returned Overseas Scholars in Shanxi Province under Grant No.20220008, and the Shanxi Province Science Foundation under Grant No.202303021221031.

\end{document}